\documentclass[fleqn,usenatbib]{mnras}

\usepackage{ulem}
\usepackage{mathptmx}
\usepackage{amssymb}
\usepackage{amsmath}

\usepackage[T1]{fontenc}

\DeclareRobustCommand{\VAN}[3]{#2}
\let\VANthebibliography\thebibliography
\def\thebibliography{\DeclareRobustCommand{\VAN}[3]{##3}\VANthebibliography}

\usepackage{graphicx}	
\usepackage{amsmath}	

\title[AGN winds in IRAS19154+2704]{AGN driven outflows in the OH absorber galaxy IRAS\,19154+2704}

\author[C. Hekatelyne et al.]{
C. Hekatelyne,$^{1,2}$\thanks{E-mail: hekatelyne.carpes@gmail.com (CH)}
Thaisa Storchi-Bergmann,$^{2}$\thanks{E-mail: thaisa@ufrgs.br (TSB)} Rogemar A. Riffel,$^{3}$\thanks{E-mail: rogemar@ufsm.br (RAR)} Preeti Kharb,$^{4}$ Claudia M. Cassanta$^{3}$
\newauthor Andrew Robinson,$^{5}$ Dinalva A. Sales$^{6}$  
\\
$^{1}$ Universidad Tecnol\'ogica, Polo Educativo Tecnol\'ogico Rivera, Ruta 5 km 496, CP 40000, Rivera, Uruguay\\
$^{2}$Departamento de Astronomia, Universidade Federal do Rio Grande do Sul, 91501-970, Porto Alegre, RS, Brazil\\
$^{3}$Departamento de F\'isica, CCNE, Universidade Federal de Santa Maria, 97105-900, Santa Maria, RS, Brazil\\
${^4}$ National Centre for Radio Astrophysics, Tata Institute of Fundamental Research, S. P. Pune University Campus, Post Bag 3,\\ Ganeshkhind, Pune 411 007, India \\
$^{5}$ School of Physics and Astronomy, Rochester Institute of Technology, 84 Lomb Memorial Drive, Rochester, NY 14623, USA \\
$^{6}$ Instituto de Matem\'atica, Estat\'istica e F\'isica, Universidade Federal do Rio Grande, Rio Grande 96203-900, Brazil \\
}

\date{Accepted XXX. Received YYY; in original form ZZZ}

\pubyear{2023}

\begin{document}
\label{firstpage}
\pagerange{\pageref{firstpage}--\pageref{lastpage}}
\maketitle

\begin{abstract}
We present a two-dimensional study of the gas distribution, excitation and kinematics of the OH absorber  galaxy IRAS\,19154+2704 using Gemini GMOS-IFU observations. Its continuum image shows a disturbed morphology indicative of a past or on-going interaction. The ionised gas emission presents two kinematic components: a narrow ($\sigma\lesssim$300\,km\, s$^{-1}$) component that may be tracing the gas orbiting in the galaxy potential and a broad ($\sigma\gtrsim$500\,km\,s$^{-1}$) component which is produced by an Active Galactic Nucleus (AGN) driven outflow, with velocities reaching $-$500\,km\,s$^{-1}$ which may exceed the escape velocity of the galaxy. The emission-line ratios and BPT diagrams confirm  that the gas excitation in the inner $\sim$2 kpc is mainly due the AGN, while in regions farther away, a contribution from star formation is observed. 
We estimate a mass outflow rate of $\dot{M}_{\rm out}=4.0\pm2.6$\,M$_\odot$\,yr$^{-1}$ at a distance of 850 pc from the nucleus. The corresponding outflow kinetic power $\dot{E}_{\rm out} = (2.5\pm1.6)\times10^{42}$\:erg\,s$^{-1}$, is only $3\times10^{-4}$ L$_{\rm bol}$ (the AGN luminosity), but the large mass-outflow rate, if kept for a $\sim$10\,Myr AGN lifecycle, will expel $\approx10^8$\:M$_\odot$ in ionised gas alone.  This is the 6th of a series of papers in which we have investigated the kinematics of ULIRGS, most of which are interacting galaxies showing OH Megamasers. IRAS19154 shows the strongest signatures of an active AGN, supporting an evolutionary scenario: interactions trigger AGN that fully appear in the most advanced stages of the interaction. 
\end{abstract}




\section{Introduction}

The process of galaxy mergers and interactions is of paramount importance in the evolution of galaxies, as it leads to modifications in their morphologies, physical and chemical properties
\citep[e.g.][]{Mihos95,Schwarzkopf00,Perez11,krabbe14,araujo23}. (Ultra)Luminous Infrared Galaxies (U)LIRGs -- with infrared luminosities of $L_{\rm IR}>10^{11}$\:L$_\odot$ --  are the result of complex processes involving the interactions and mergers of galaxies  and symbolize a key phase in the evolutionary path of galaxies, during which the gravitational interactions from mergers exert tidal torques, channeling gaseous material into the heart of the galaxy \citep{Soifer87,sanders88}. This process can trigger intense star formation (SF) and the fuelling an Active Galactic Nucleus  \citep[e.g.][]{Tissera02,Petric18,SB-SM19}.

ULIRGs also appear to be the primary hosts of OH megamaser (OHM) emitters, observed as a result of stimulated emission phenomena primarily emitting in the 1667 and 1665 MHz lines, with luminosities ranging from 10$^2$ to 10$^4$ L$_\odot$ \citep{Darling2000,Darling2002,Lo2005}. OHM galaxies may represent a short-lived transition phase during which dense concentrations of molecular gas trigger intense star formation episodes and the eruption of an AGN  \citep{sanders88,Barnes92,Hopkins2006,Haan2011}.

To better understand the relationship between ULIRGs, AGN SF, and OHM galaxies, we have an ongoing project dedicated to investigating the gas emission structure of a sample of roughly 70 galaxies selected from \citet{Darling2000}, most of them with OHM emission. We use Hubble Space Telescope (HST) images, Very Large Array (VLA) observations and Integral Field Spectroscopy (IFS) data to characterize the gas emission-structure and kinematics in the inner few kiloparsec of these galaxies. The overal goal of our project is relating the merger stage and OH maser properties to quantitative measurements of AGN and Starburst activity. We have so far focused on a sub-sample of 15 galaxies for which we have HST broad band and ramp filter H$\alpha$+[N{\sc ii}] observations \citep{Dinalva2015,Heka2018a,Heka2018b,Dinalva2019,Heka2020}. This HST sample includes a range of merger stages and covers the redshift range $z = 0.009 - 0.26$, and a range of OH maser luminosities and OH line FWHM that are representative of the population of known OHMGs. 

Our main results to date are summarized as follows. \citet{Dinalva2015} concluded that the late-stage merger IRAS16399-0937 hosts an embedded AGN in one nucleus that is probably the source of the OHM emission observed in this system, while the second nucleus is starburst dominated. For IRAS\,F23199+0123, \citet{Heka2018a} revealed an obscured type 2 AGN and reported a new OHM source related to the AGN. Also using multiwavelength observations, \citet{Heka2018b} reported a possible embedded AGN in the midst of
a circumnuclear ring of star formation in IRAS\,03056+2034. \citet{Dinalva2019} identified two kinematic components in IRAS\,17526+3253 and concluded that the main source of ionization of this galaxy is young stars, although the presence of an AGN could not be discarded. And finally, \citet{Heka2020} revealed that a faint AGN in IRAS\,11506-3851 could be the origin of previously reported outflows in neutral and molecular gas. In summary, in our previous studies we have found five OHMs for which the GMOS-IFU data have resolved previously unknown AGNs and -- although the number of objects is still small -- our results support the hypothesis that OHM galaxies are hosts of AGNs that are being triggered by recent accretion of matter to their central Supermassive Black Hole (SMBH).

In this paper, we focus on a source with host properties similar to those of the galaxies previously studied, but with the OH feature observed in absorption. We present a detailed study of the ULIRG IRAS\,19154+2704 (hereafter IRAS\,19154), which has a redshift z=0.0994 \citep{Nakanishi1997}, far infrared (FIR) luminosity L$_{FIR}$ 10$^{10.66}$ L$_{\odot}$ and is classified as an OH absorber by \citet{Darling2000}. These authors reported that the 1665 MHz absorption line is strong and has a velocity in agreement with the autocorrelation function secondary peak and line separation predictions. Also, as reported by these authors, the host presents a fairly strong 1.4 GHz continuum with flux density of 64 mJy. Assuming the Hubble constant to be H$_{0}$=70.5 km s$^{-1}$ Mpc$^{-1}$, and considering a distance of 436 Mpc, the adopted scale at the galaxy is 2.114 kpc arcsec $^{-1}$.

We use a combination of Gemini IFU data and HST images to map and study the gas distribution, kinematics and excitation in IRAS19154. In addition we present a 1.4~GHz continuum image obtained with the VLA.  In section 2 we provide information on the observations and data reduction and in section 3 we describe the emission-line profile fitting and resulting gas distribution, excitation and kinematic results. In Section 4 we discuss the gas kinematics and the nature of the gas excitation and finally we present our conclusions in Section 5.

\section{Observations and data reduction}\subsection{GMOS-IFU data}

The spectroscopic data of IRAS19154 were obtained at the Gemini North Telescope with the Gemini Multi Object Spectrograph \citep[GMOS;][]{hook04} operating in the Integral Field Unit (IFU) \citep{allington-smith02} mode. The GMOS-IFU data were obtained on the night of 2017 September 11, split into nine observations of 1180 seconds each, with the IFU operating in one slit mode using the grating B600-G5323 and the IFU-R mask. Moreover we obtained spectra centered at wavelengths 6200\AA\ and 6300\AA\ in order to allow recovery of the signal within the detector gaps. 

Using this configuration, we obtained a field of view of 3.5\farcs $\times$ 5\farcs0 and recovered information from a spectral range that includes strong emission lines from H$\beta$ to [S\,{\sc ii}]$\lambda$6731\AA. The data reduction followed the standard procedures for spectroscopic data and was performed using the GEMINI IRAF package. The basic steps of the reduction are the bias level subtraction, trimming and flat-fielding, wavelength calibration, sky subtraction and relative flux calibration, using observations of the standard star LTT\,9239 performed at the same night of the observations of the galaxy. 

As the final step, we median-combined the individual data cubes  using the position of the peak of the continuum emission to perform the astrometry among the individual data cubes. We derived a seeing of 0.65 arcsec from the measurement of the full width at half maximum (FWHM) of the flux distribution of field stars available in the acquisition image obtained just before the observations of IRAS19154. The combined data cube covers the inner 3\farcs4$\times$5\farcs0 of the galaxy at an angular sampling of 0\farcs05$\times$0\farcs05.

\begin{figure*}
	\includegraphics[width=0.9\textwidth]{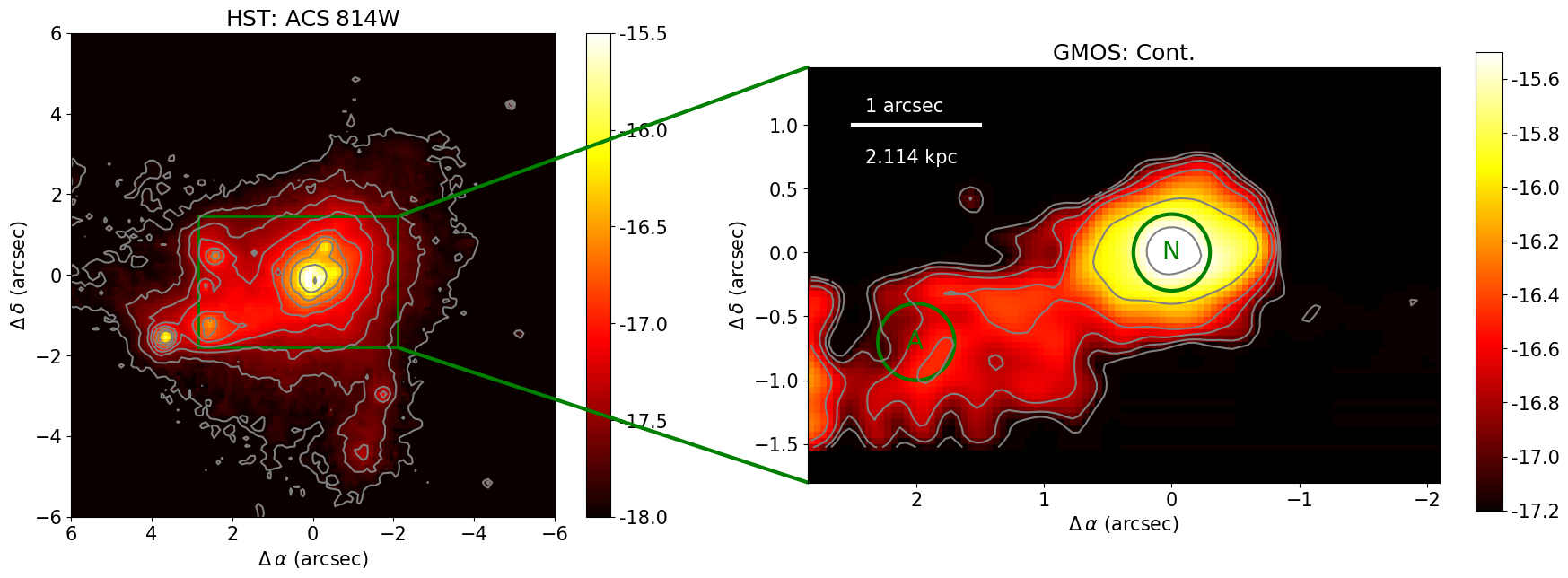}
	\includegraphics[width=0.9\textwidth]{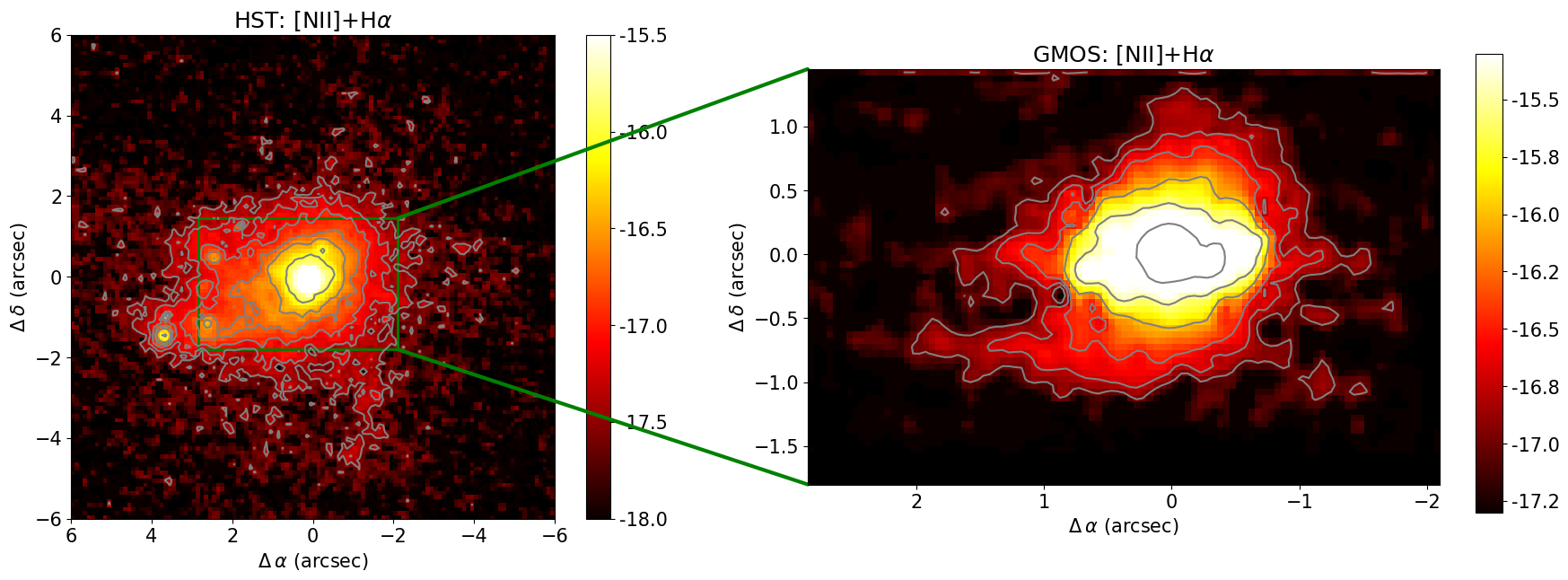}
 	\includegraphics[width=0.9\textwidth]{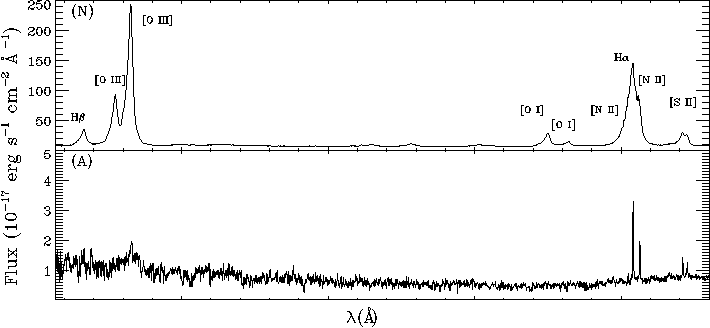}
    \caption{Top and middle: HST and GMOS-IFU images of IRAS\,19154.
   The top and middle left panels correspond to the HST/ACS F814W - i band image of the galaxy and the HST H$\alpha+$[N\,{\sc ii}]$\lambda\lambda6548,84$ narrow-band image, respectively. The top and middle right panels show the corresponding GMOS-IFU continuum image within a spectral window 5500--6000\AA (top) and another window covering the  H$\alpha$+[N\,{\sc ii}]$\lambda\lambda6548,84$ emission lines (bottom), respectively. The color bars show the continuum in logarithmic units of erg\:s$^{-1}$\:cm$^{-2}$\:\AA\:$^{-1}$\:arcsec$^{-2}$ and the H$\alpha$+[N\,{\sc ii}]$\lambda\lambda6548,84$ fluxes in logarithmic units of erg\:s$^{-1}$\:cm$^{-2}$\: $^{-1}$\:arcsec$^{-1}$. The gray contours represent levels of flux from each image, with the aim of enhancing fainter structures. The bottom plots show two representative spectra of the different regions probed by our observations: the nucleus (N), that shows broad components in the emission lines, and a region with no evidence of broad component (A). These regions are identified in the GMOS continuum image.}  
    \label{fig:HST}
\end{figure*}

Figure\,\ref{fig:HST} shows, in the two upper rows, HST (left) and GMOS (right) images of IRAS19154, in the continuum (top) and in the H$\alpha$+[N\,{\sc ii}] emission lines (bottom). The two bottom rows,  show two representative GMOS spectra: one centered on the position of the nucleus (labeled N) -- defined as the location of the continuum peak -- and the other from a location at $\sim$2 arcsec east of the nucleus (labeled A). These spectra are integrated within an aperture radius of 0.3 arcsec. The strongest emission lines are identified in the nuclear spectrum and the comparison between the two spectra clearly reveals broader emission lines at the nucleus, compared to extranuclear regions.

\subsection{GMOS-IFU Emission-line profile fitting}

In order to obtain measurements of the gas properties in IRAS\,19154 we used the {\sc ifscube} python package \citep{Ruschel-Dutra2020,Ruschel-Dutra2021} to model the emission-line profiles by Gaussian curves.  We first use {\sc ifscube} to model the emission lines at the nucleus. By visual inspection of the spectra, we notice that a single Gaussian component can not reproduce the observed line profiles, as they usually present a broad wing in the nuclear region and thus, we include two Gaussian components per emission line. These components will be referred to as {\it broad} and {\it narrow} components, hereafter. We use the {\it cubefit} {\sc ifscube} module, providing as initial guesses for the centroid velocity and velocity dispersion ($\sigma$) the values obtained from the fits of the line profiles using the {\sc iraf.splot} task, while the fluxes at the peak of the line profiles estimated directly with the {\sc ifscube} code are used as guesses for the Gaussian amplitudes. The fit of the emission-line profiles is performed in a continuum-subracted cube, where the continuum emission is represented by a 5$^{\rm th}$ order polynomial.
These fits are performed from the nuclear spaxel outwards, following a spiral pattern. For each spaxel, initial guesses for the free parameters are provided as the values derived from successful fits of spaxels located at distances smaller than 0.15 arcsec.

The following emission lines are fitted simultaneously: H$\beta$, [O\,{\sc iii}] $\lambda\lambda$4959,5007, [O\, {\sc i}]$\lambda\lambda$6300,6364, [N\,{\sc ii}] $\lambda\lambda$6548,6583, H$\alpha$ and   [S\,{\sc ii}] $\lambda\lambda$6717,6731.  The [O\,{\sc iii}]$\lambda$5007/[O\,{\sc iii}]$\lambda$4959 and [N\,{\sc ii}]$\lambda$6583/[N\,{\sc ii}]$\lambda$6548 line ratios are kept fixed to their theoretical values of 2.98 and 3.06, respectively \citep{Osterbrock2006}. The centroid velocity and velocity dispersion of emission lines from  the same parent ions ([O\,{\sc iii}], [N\,{\sc ii}], [S\,{\sc ii}] and H\,{\sc i}) are kept tied, separately for the broad and narrow components.

\begin{figure}
\centering
	\includegraphics[width=0.8\columnwidth]{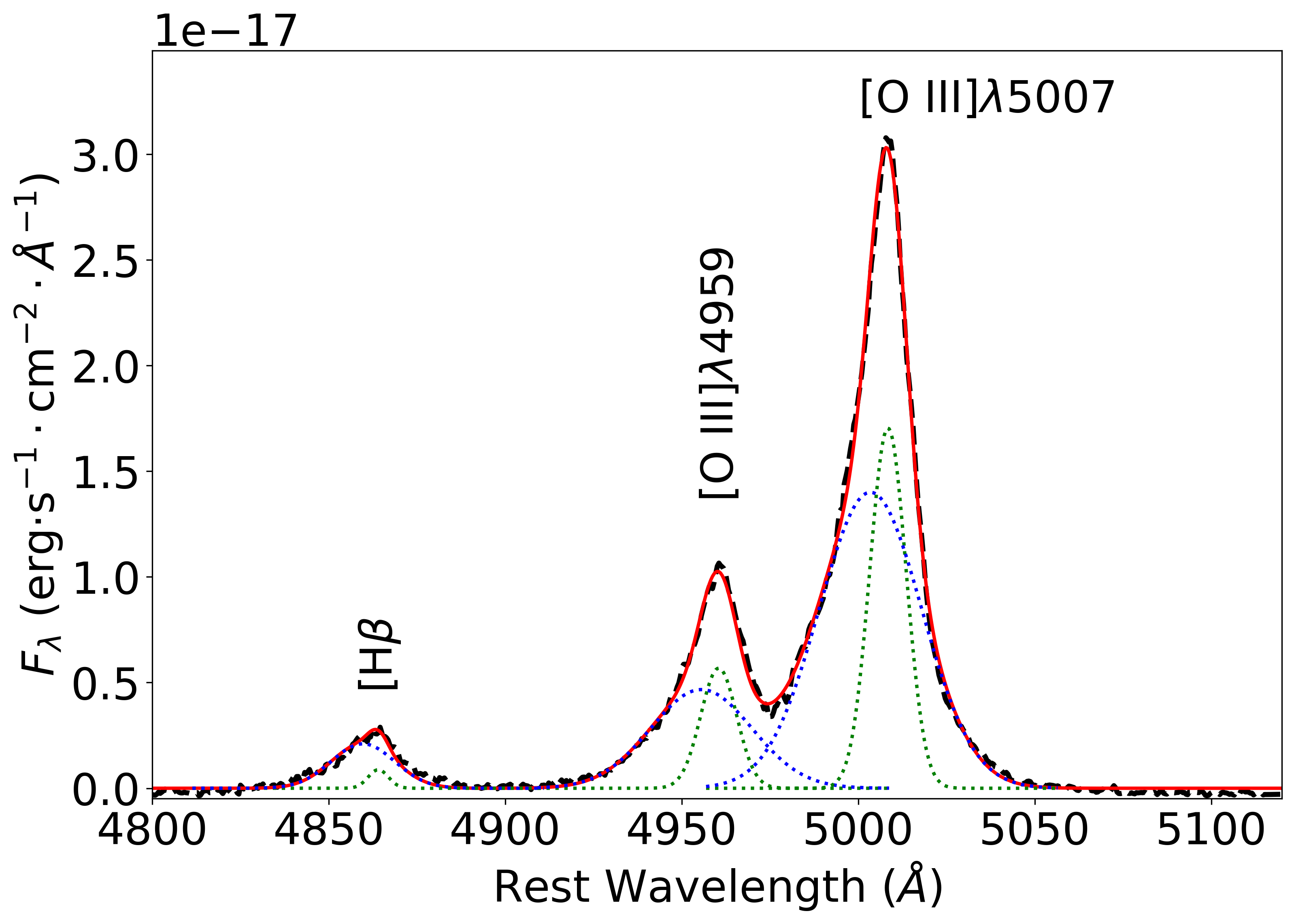}
	\includegraphics[width=0.8\columnwidth]{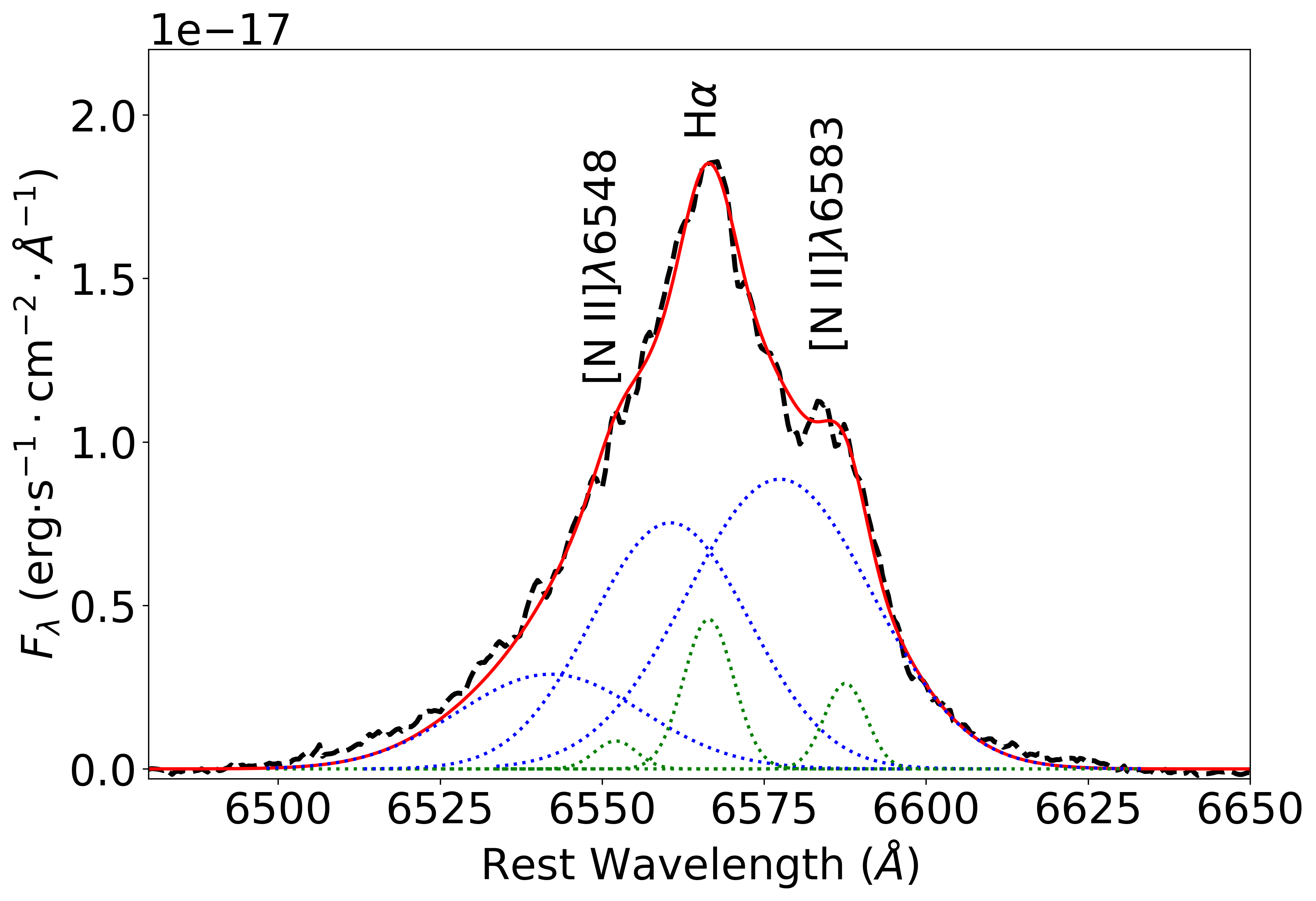}
	\includegraphics[width=0.8\columnwidth]{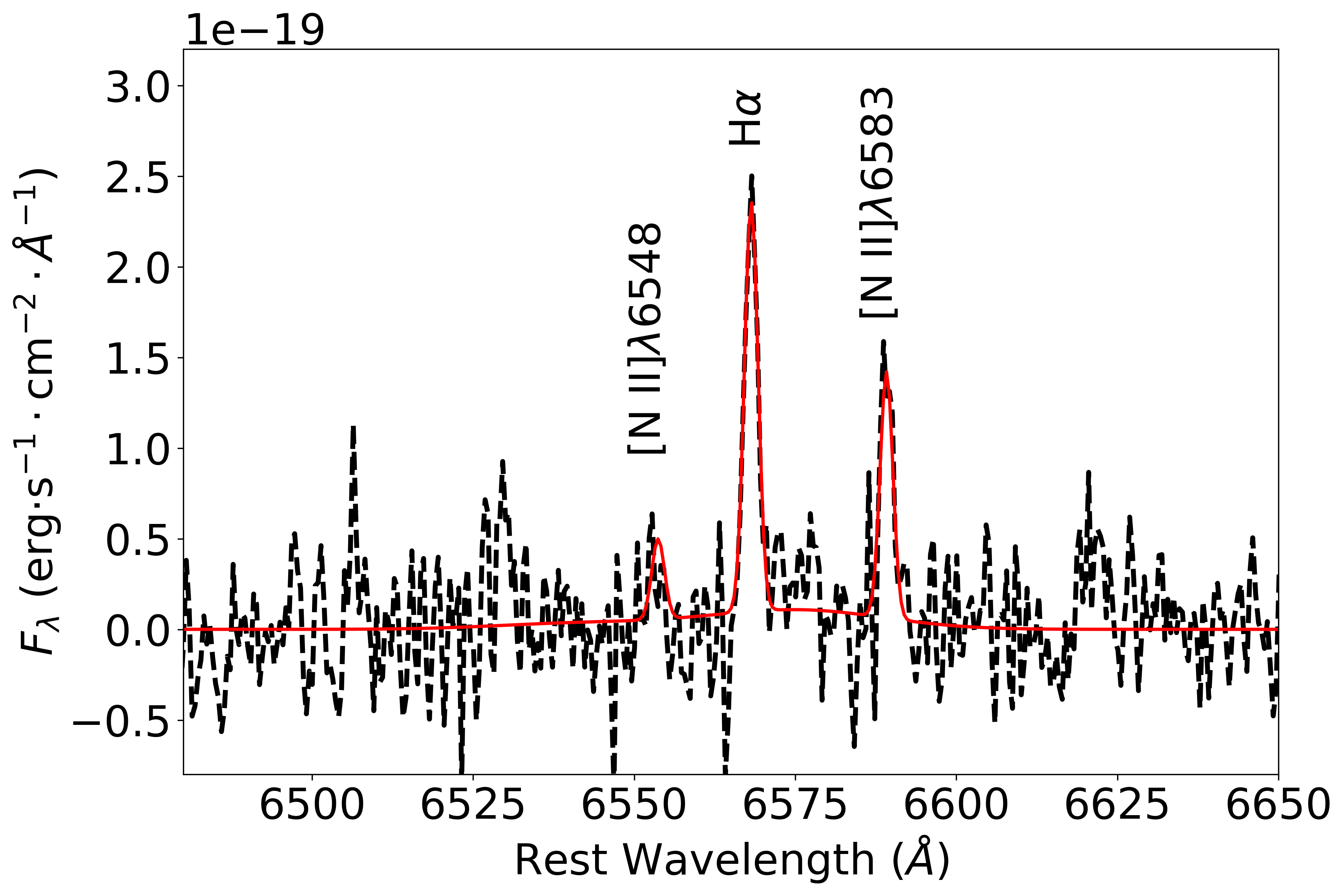}
    \caption{Examples of fits of the [O{\sc iii}]+H$\beta$ (top panel) and [N\,{\sc ii}]+H$\alpha$ (middle panel) emission line profiles for the nuclear spectrum and for a position at (0\farcs75\:S, 2\farcs0\:E) of the nucleus (bottom panel). The nuclear spectrum was fitted by two Gaussian curves per emission-line, while a single component per line reproduces the extra-nuclear profiles. In each panel, the observed spectrum is shown a dashed black line and the model is shown in red. The individual Gaussian components for the nuclear profiles are shown as dotted lines in blue for the broad component and in green for the narrow component. In both panels, the spectra are continuum subtracted.}
    \label{fig:ajuste}
\end{figure}

Figure\,\ref{fig:ajuste} shows examples of emission-line fits for the H$\beta$+[O {\sc iii}]$\lambda\lambda$5007,4959  and H$\alpha+$[N\,{\sc ii}]$\lambda\lambda6548,6583$ complexes for the nuclear spaxel in the top and middle panels, respectively. As can be noticed, both {\it narrow} and {\it broad} Gaussian components are required to fit the emission-line profiles at the nuclear region of IRAS\,19154. The bottom panel shows  the fit of the H$\alpha+$[N\,{\sc ii}]$\lambda\lambda6548,6583$  profiles extracted at the region labelled as $A$ in Fig.\,\ref{fig:HST}, separated from the nucleus by $\sim$2 arcsec, for which the emission-line profiles are well represented by a single Gaussian component, narrower than that of the {\it narrow} component of the nuclear spectrum.

\subsection{HST data}

The HST images were obtained as part of a snapshot survey program designed to observe a sample of OHMGs. The observations were performed using the Advanced Camera for Surveys (ACS) with broad (F814W), narrow (FR656N) and medium-band (FR914M) filters. The data reduction followed the standard procedure for HST images and more details can be seen in \citet{Dinalva2015,Dinalva2019}.

We have used F814W filter in order to examine the host galaxy structure. The ramp filter images were acquired for the purpose of investigating the distribution of ionized gas. The central wavelengths were configured to encompass H$\alpha$ in the narrow-band filter and the adjacent continuum for continuum subtraction in the medium-band filter. The bandpass of the FR656N ramp filter encompasses FR914M as well as the H$\alpha$ and the [N{\sc ii}]$\lambda6548,83$ lines. The total integration times in the F814W, FR914M and narrow H$\alpha$ FR656N filters were 600, 200, and 600 seconds, respectively.

The pipeline image products were further processed with the IRAF \footnote{IRAF is  distributed by the National Optical Astronomy Observatory, which is operated by the Association  
of Universities for Research in Astronomy (AURA), Inc., under cooperative agreement with   the National Science Foundation.} task $lacos_{im}$ \citep{lacos} to remove cosmic rays. The continuum-free H$\alpha+$[N{\sc ii}] image for IRAS19154 was generated using a well-established procedure. This method initially calculates the count rates for several foreground stars in both the medium (FR914M) and narrow-band (FR656N) ramp filter images \citep[see][]{Hoopes99,Rossa00,Rossa03,Dinalva2015}. 
The subsequent step involved determining the mean scaling factor based on the count rate ratios. This factor was then applied to the medium-band FR914M image, which was subsequently subtracted from the narrow-band FR656N image.
The continuum-subtracted H$\alpha+$[N{\sc ii}] image was examined carefully to ensure that there were no significant residuals at the locations of the foreground stars in the image. The uncertainties associated with this method are usually $\sim 5-10$\% \citep[see][]{Hoopes99,Rossa00,Rossa03}.

\subsection{VLA Radio Continuum data}

\begin{figure}
\centerline{
	\includegraphics[width=9cm,trim=20 150 10 150]{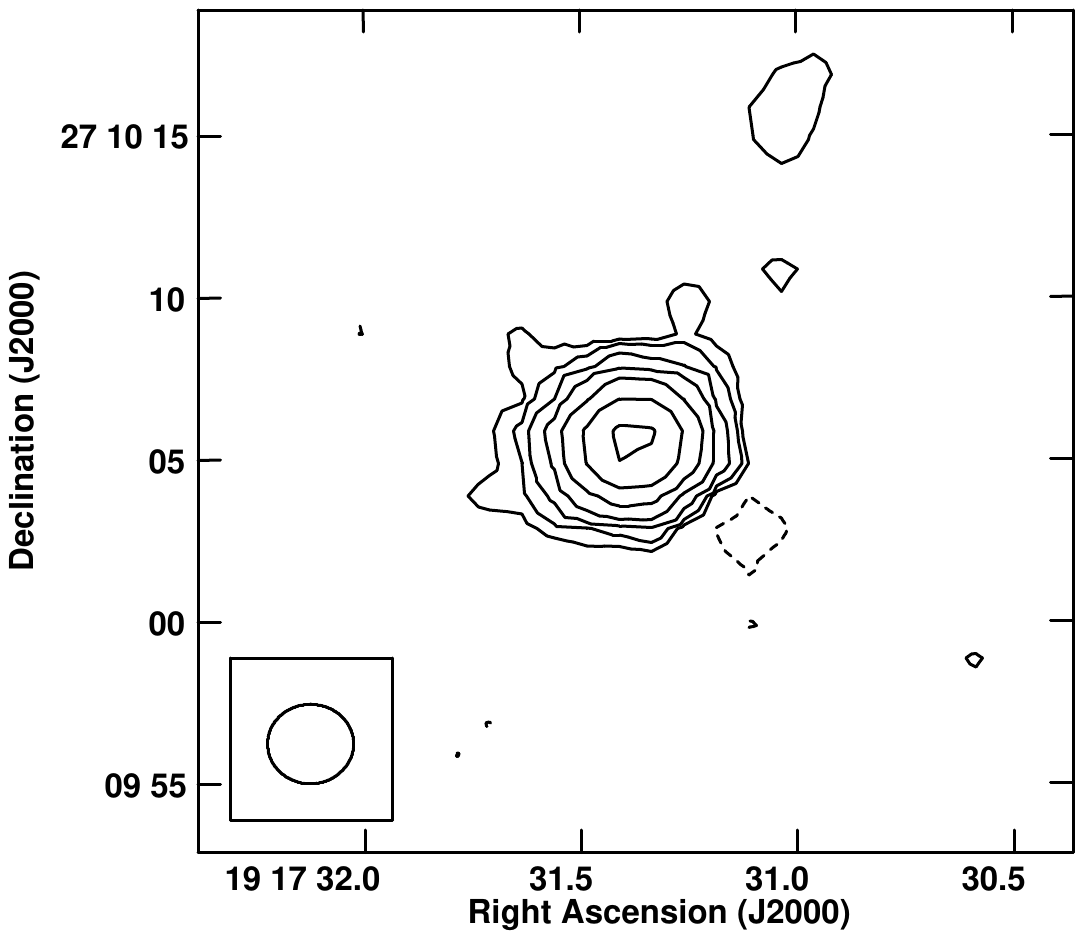}}
    \caption{The VLASS image of IRAS\,19154 at 3.0~GHz. A single radio component is observed. The contour levels are in percentage of the peak surface brightness ($I_p = 26.1$~mJy~beam$^{-1}$) and increase in steps of two with the lowest contour level being at $\pm1.4$\% of $I_p$. The beam is $2.66\arcsec \times 2.45\arcsec$ at a position angle of $-89.5\degr$.}
    \label{fig:vla}
\end{figure}

 IRAS\,19154 has been observed as part of the Karl G. Jansky Very Large Array Sky Survey \citep[VLASS][]{Lacy2020} at 3.0 GHz. A point source is observed with a peak intensity of $29.5\pm0.1$~mJy~beam$^{-1}$ and an integrated flux density of $\sim30.7\pm0.2$~mJy (see Fig.\,\ref{fig:vla}). These estimates were obtained using the {\tt AIPS} Gaussian-fitting task {\tt JMFIT}. The similarity in the peak and integrated values indicates that the radio source is compact. The VLASS epoch~1.2 image from May 17, 2019, has a synthesized beam size of $2.66\arcsec \times 2.45\arcsec$ at a position angle of $-89.5\degr$. The $rms$ noise in the image is 0.14~mJy~beam$^{-1}$. {\tt JMFIT} further indicates that the beam-deconvolved size of the radio source is only $\sim0.8\arcsec$. 

 IRAS\,19154 has also been observed as part of the NRAO VLA Sky Survey (NVSS) at 1.4~GHz on May 16, 1995 \citep{Condon1998}, where a point source is detected with a peak intensity of $61.6\pm0.5$~mJy~beam$^{-1}$ and an integrated flux density of $\sim64.9\pm0.9$~mJy. The NVSS image has a beam-size of $45\arcsec \times 45\arcsec$, and the $rms$ noise in the image is 0.43~mJy~beam$^{-1}$. From a comparison of the NVSS and VLASS images it is evident that $\gtrsim50$\% of the radio emission emerges from a $\sim0.8\arcsec$-scale region. The peak intensities in the NVSS image and in the VLASS image after convolution with the NVSS beam of $45\arcsec$ indicates a $1.4-3.0$~GHz spectral index of $-0.85\pm0.15$ ($S_\nu\propto\nu^\alpha$). This value is consistent with synchrotron emission and inconsistent with free-free emission. Moreover, the brightness temperature of the radio component \citep[see][]{Ulvestad2005} is $\sim1.0\times10^4$~K, which is similar to that observed in arcsecond-scale observations of other radio-quiet AGN \citep{Kukula1998,Berton2018}.

\section{Results}
\subsection{HST images}
The upper-left and central-left panels of Figure \ref{fig:HST} present the ACS/HST F814W i-band and the narrow-band H$\alpha+$[N\,{\sc ii}]$\lambda\lambda6548,6584$ images of IRAS\,19154, respectively. The right panels show the corresponding GMOS-IFU images.

 The $i$-band image shows the emission peak at the nucleus. This emission structure is more elongated towards the east, with three unresolved knots of enhanced emission at $\approx$1--4$^{\prime\prime}$ from the nucleus, in the elongated structure. Moreover there is a faint elongation to the south of the nucleus. The GMOS continuum image,  shown in the top-right panel of Figure \ref{fig:HST}, was obtained by computing the mean fluxes between 5500\AA\ and 6000\AA, a spectral region free of strong emission lines. It shows a similar elongated structure to the east as observed in the HST image, but the emission knots seen in the HST images are less evident. By convolving the HST image with a Gaussian kernel with a sigma corresponding to the seeing of the GMOS observations, the emission knots disappear. This indicates that the knots of emission are not seen in the Gemini data due to the poorer angular resolution of GMOS compared to HST.  

The HST continuum-free H$\alpha+$[N\,{\sc ii}]$\lambda\lambda6548,6584$ narrow-band image, presented in the central left panel of Fig~\ref{fig:HST}, shows a similar flux distribution to that of the continuum $i$-band image (top-left panel), showing similar centrally peaked emission and knots to the east of the nucleus.  There is also some emission in the faint elongation to the south. A similar structure is observed in the GMOS {\it pseudo-narrow band} image, obtained by integrating the fluxes in the H$\alpha+$[N\,{\sc ii}]$\lambda\lambda6548,6584$ region and subtracting the continuum from adjacent regions.

\subsection{GMOS -- Emission-line flux distributions and kinematics}

\begin{figure*}
	\includegraphics[width=\textwidth]{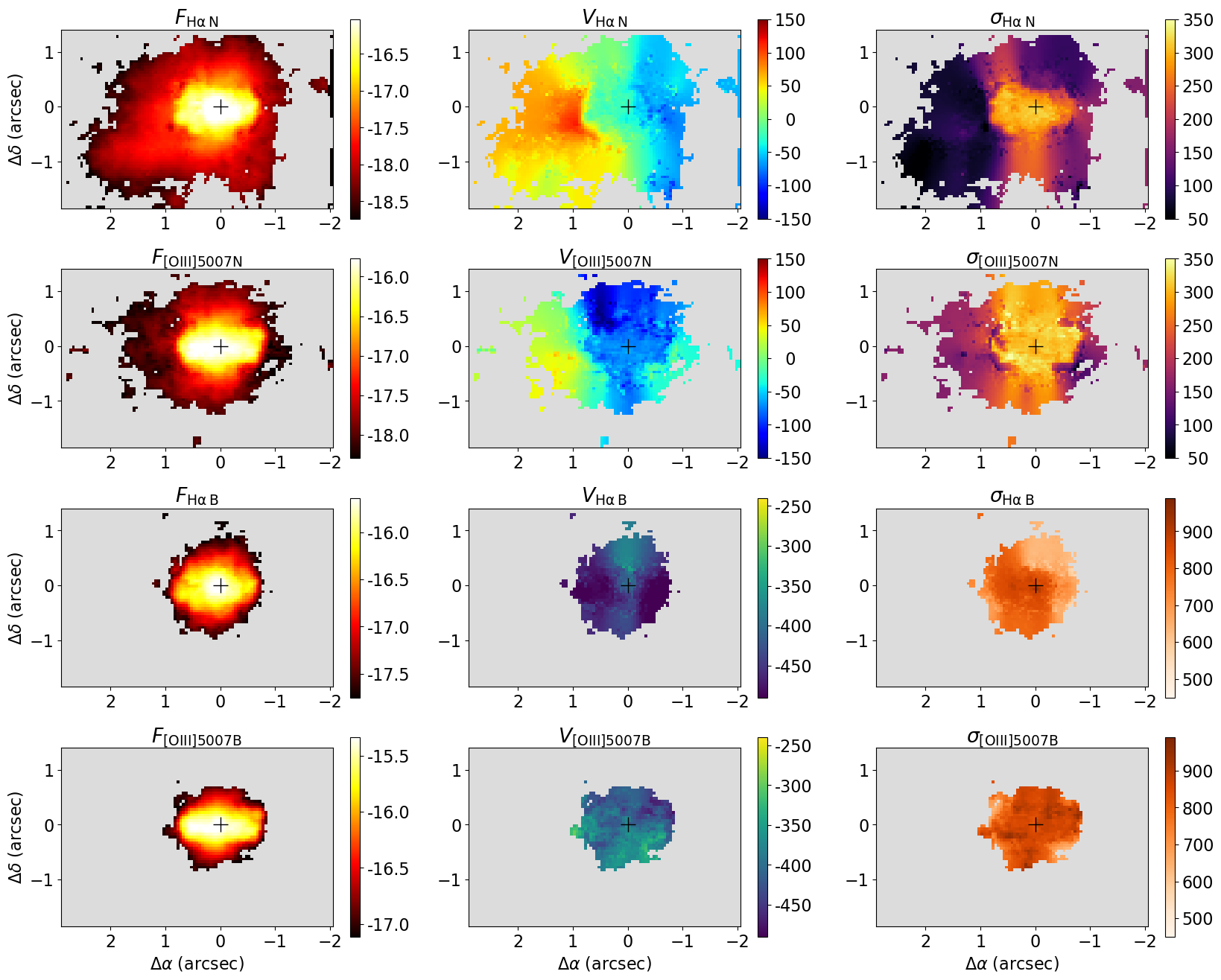}
    \caption{Flux distributions and kinematics for the [O\,{\sc iii}]$\lambda$5007 and H$\alpha$ emission lines of IRAS19154, based on the GMOS data. The left column displays the flux distributions, for the H$\alpha$ narrow component,  [O\,{\sc iii}]$\lambda$5007 narrow component, H$\alpha$ broad component, and [O\,{\sc iii}]$\lambda$5007 broad component, from top to bottom. The central column shows the corresponding line of sight velocity fields and the right column presents the velocity dispersion maps. In all panels the central crosses mark the position of the nucleus and grey regions indicate masked locations  where signal-to-noise was not high enough to allow measurements or locations with no detection of line emission. The color bars show the fluxes in logarithmic units of erg\,s$^{-1}$\,cm$^{-2}$\,spaxel$^{-1}$ and the velocities and velocity dispersions in km\,s$^{-1}$. The velocity fields are shown after correction for the systemic velocity of the galaxy and the $\sigma$ maps are corrected for instrumental broadening.}
    \label{fig:gmos}
\end{figure*}

\begin{figure*}
	\includegraphics[width=\textwidth]{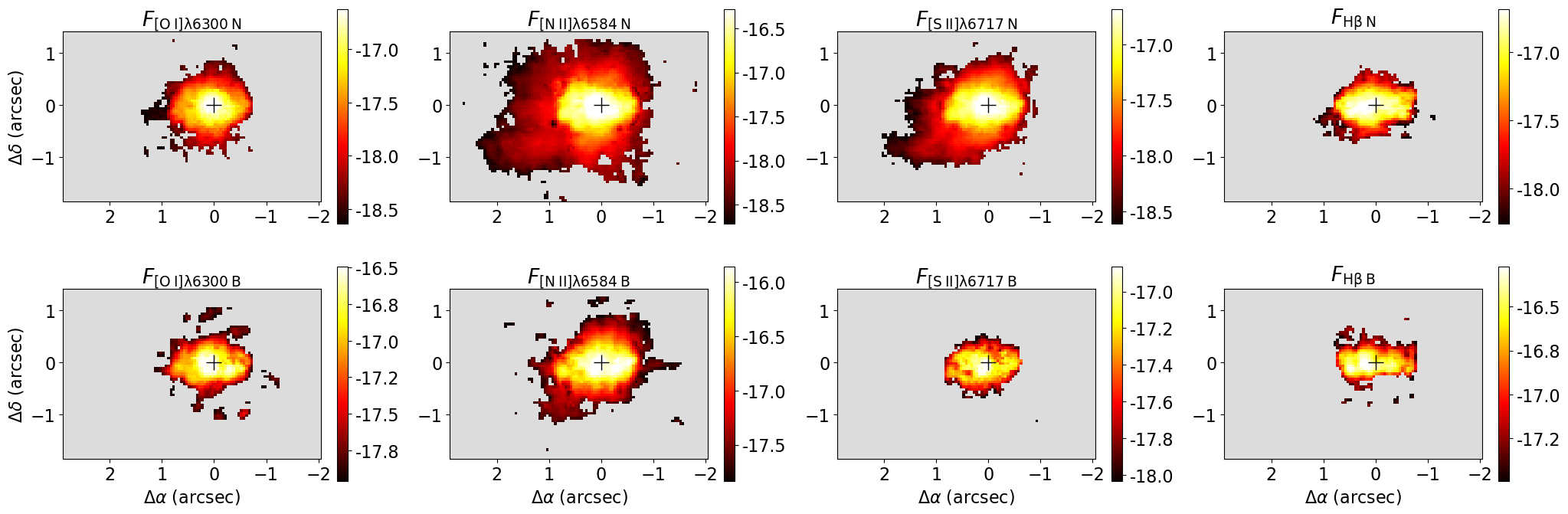}
    \caption{ Emission-line flux distributions for the narrow (top) and broad (bottom) components, based on GMOS data. The maps for [O\,{\sc i}]$\lambda$6300, [N\,{\sc ii}]$\lambda$6584, [S\,{\sc ii}]$\lambda$6717 and H$\beta$ are shown from left to right.  In all panels the central crosses mark the position of the nucleus and grey regions indicate masked locations  where signal-to-noise was not high enough to allow measurements or locations with no detection of line emission. The color bars show the fluxes in logarithmic units of erg\,s$^{-1}$\,cm$^{-2}$\,spaxel$^{-1}$.}
    \label{fig:fluxes}
\end{figure*}

The left panels of Figure~\ref{fig:gmos} present the flux distributions for the H$\alpha$ and [O\,{\sc iii}]$\lambda$5007 emission lines. The top two panels show the maps for the narrow components and the bottom two present the maps for the broad components. The flux maps for [O\,{\sc i}]$\lambda$6300, [N\,{\sc ii}]$\lambda$6584, [S\,{\sc ii}]$\lambda$6717, and H$\beta$ are shown in Figure~\ref{fig:fluxes}, with the results for the narrow components shown in the top panels and those for the broad components in the bottom panels.

For all emission lines, the peak flux is observed at the nucleus of the galaxy. The emission of the broad line components is restricted to an elongated structure oriented along the east-west direction, extending up to $\approx$\,1$^{\prime\prime}$ (2.11 kpc at the galaxy) of the nucleus. The maps for the narrow components exhibit a similar emission structure at the highest flux levels, but narrow emission lines are also observed farther away from the nucleus.
The most extended emission is observed for H$\alpha$ with elongated structures seen to the south-east and to the south, similarly to the larger scale emission structures observed in the HST narrow band image (Fig.~\ref{fig:HST}). A similar structure, but with decreasing extent is followed by [N\,{\sc ii}]$\lambda$6583, [O\,{\sc iii}]$\lambda$5007,  and [S\,{\sc ii}]$\lambda$6717, covering up to 3$^{\prime\prime}$ (6.3 kpc at the galaxy) of the FoV, while for [O\,{\sc i}]$\lambda$6300 and H$\beta$ the emission is restricted to the central $\sim$1$^{\prime\prime}$ radius.

 The central panels of Fig.~\ref{fig:gmos} present the line of sight velocity fields for the H$\alpha$ and [O\,{\sc iii}]$\lambda$5007 emission lines. The top two panels display the results for the narrow components, while the bottom two panels present the velocity fields of the broad components. The color bars show the velocities in units of km s$^{-1}$.  In all panels the systemic velocity of the galaxy,  of 29975 km\,s$^{-1}$, obtained by modeling the velocity field is already subtracted.  We do not present the maps for the other emission lines because they resemble those of H$\alpha$.

 The velocity fields of the narrow components present a velocity gradient of $\pm$\,100\,km\,s$^{-1}$ 
with redshifts to the southeast and blueshifts to the northwest. The Halpha velocity field (narrow component) is dominated by a rotation pattern; a similar pattern is followed by the other low-ionization lines. The narrow component of [OIII]5007, on the other hand,  shows evidence of more disturbed kinematics. In particular, blueshifts of up to 200 km\,s$^{-1}$ are observed to the north of the nucleus. 

The velocity maps for the broad components present only blueshifts for all emission lines, with much higher velocity values than observed for the narrow components, reaching up to $-500$\,km\,s$^{-1}$.

 The right panels of Fig.~\ref{fig:gmos} present the velocity dispersion ($\sigma$) maps (corrected for the instrumental broadening) for the narrow (top two panels) and broad (bottom two panels) components of the H$\alpha$ and [O\,{\sc iii}]$\lambda$5007 emission lines.  The narrow component presents $\sigma$ values ranging from 50 to 300\,km\,s$^{-1}$ with the highest values  observed within the inner arcsecond (2.11 kpc) for all the emission lines. Overall higher values are seen for [O\,{\sc iii}]$\lambda$5007, as compared to lower ionization emission lines. The lowest $\sigma$ values, of $\sim$50\,km\,s$^{-1}$, are seen for H$\alpha$, and are mostly observed to the east at distances greater than 1 arcsec from the nucleus. The $\sigma$ maps for the broad component present values reaching up to 900\,km\,s$^{-1}$.

\subsection{GMOS -- Emission-line ratios and diagnostic diagrams}

\begin{figure*}
	\includegraphics[width=0.29\textwidth]{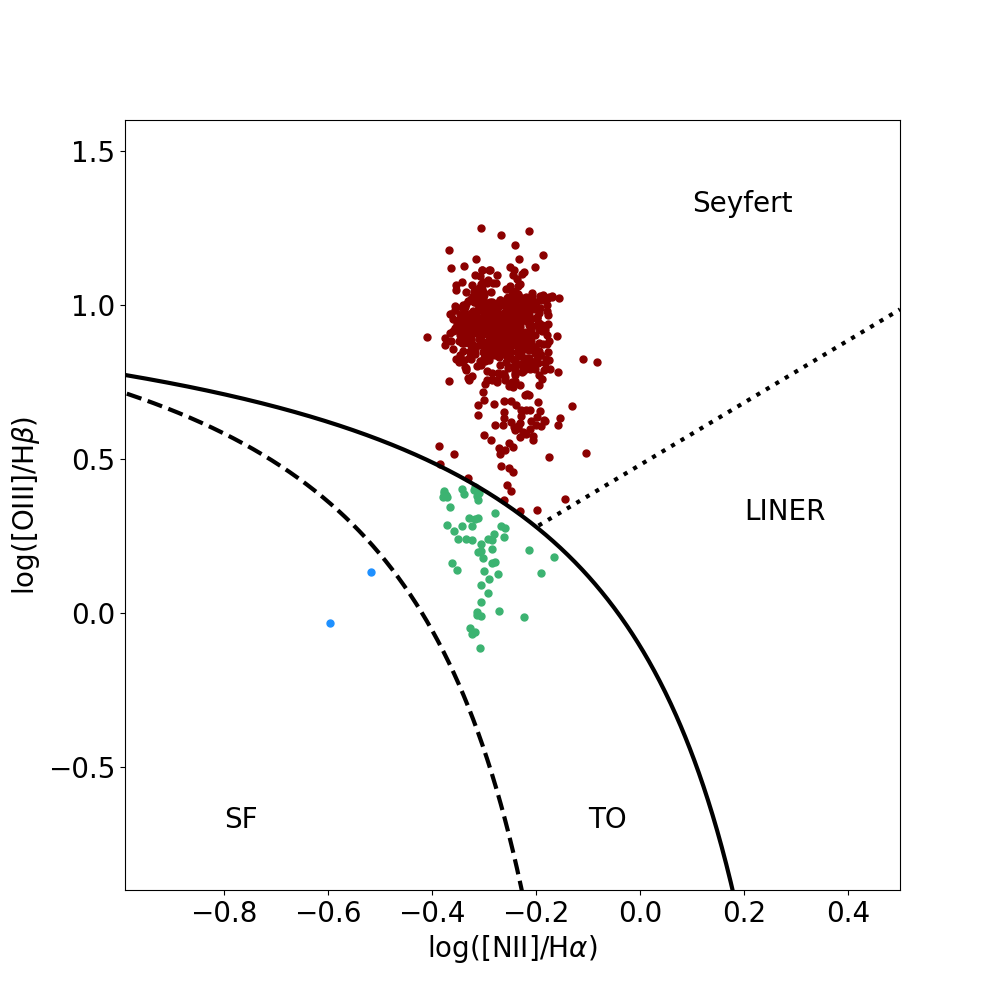}
	\includegraphics[width=0.7\textwidth]{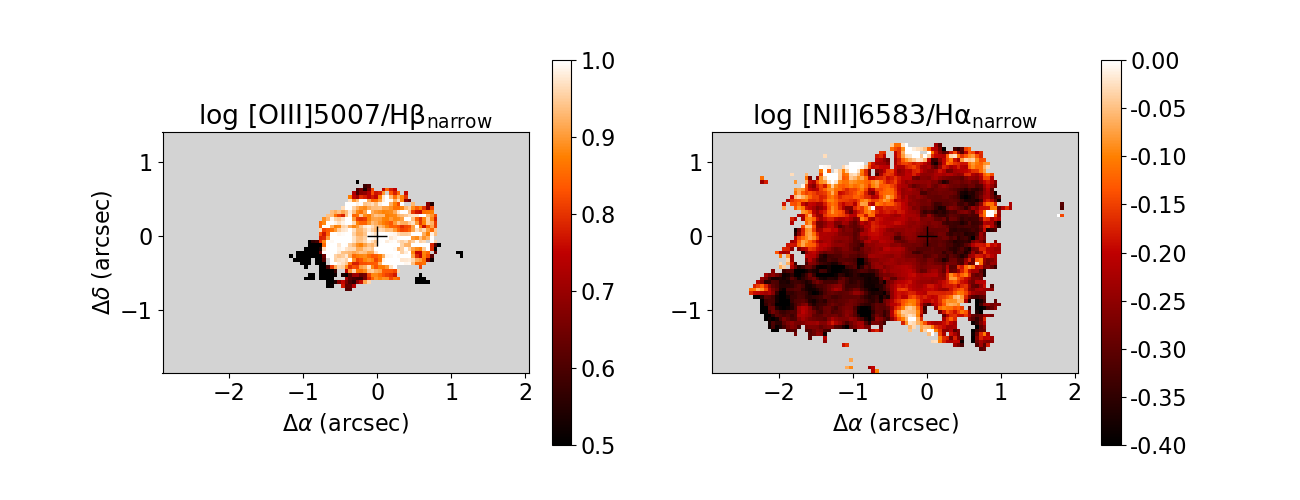}
 	\includegraphics[width=0.29\textwidth]{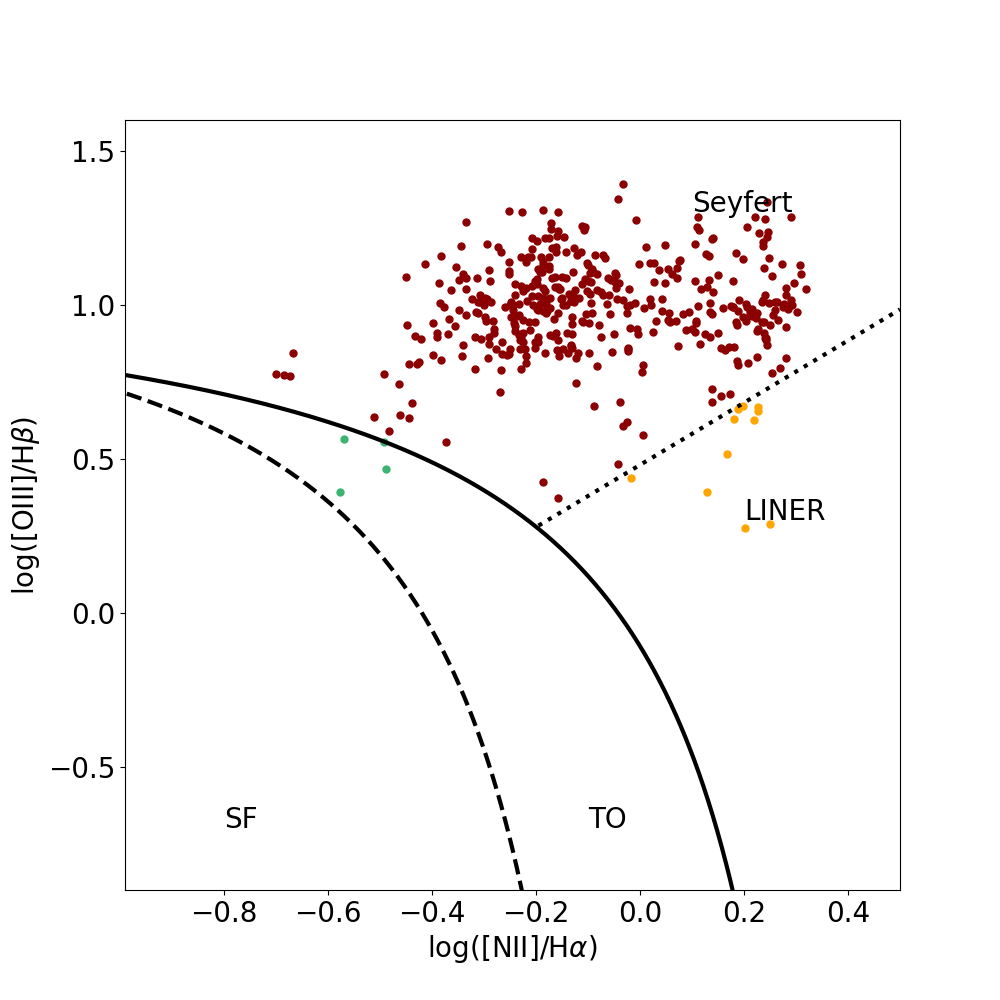}
	\includegraphics[width=0.7\textwidth]{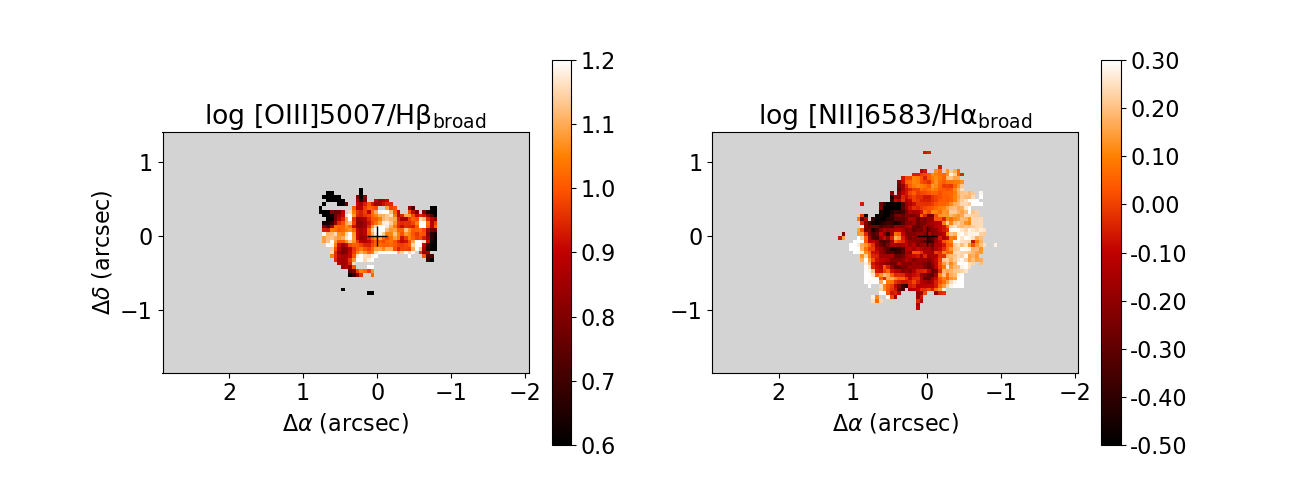}
    \caption{Top panels: log [O\,{\sc iii}]$\lambda$5007/H$\beta$ versus log [N\, {\sc ii}]/ H$\alpha$ diagnostic diagram of IRAS19154 obtained from its narrow components.  log [O\, {\sc iii}]/H$\beta$ (middle) and log [N\, {\sc ii}]$\lambda$6583/ H$\alpha$ line ratio maps (right). The dashed, continuous and dotted curves represent the borderlines from  \citet{Kauffmann2003}, \citet{kewley2006} and \citet{cid10}, respectively. Bottom panels: Same as top panels, but using the fluxes of the broad components of the emission lines. }
    \label{fig:bpts}
\end{figure*}

 The top panels of Figure\,\ref{fig:bpts} present the [O\,{\sc iii}]$\lambda$5007/H$\beta$ versus [N\, {\sc ii}]$\lambda$6584/H$\alpha$ diagnostic diagram \citep[BPT diagram;][]{baldwin81}, plotting each spaxel of the GMOS-IFU data obtained from the narrow components of the emission lines (left). The corresponding line-ratio maps [O\,{\sc iii}]$\lambda$5007/H$\beta$ (middle) and [N\,{\sc ii}]$\lambda$6584/H$\alpha$ (right) are also shown. The combination of these line ratio maps and diagnostic diagram allows us to investigate the nature of the gas excitation at each spaxel as being due to AGN or Starburst radiation or if it is due to a  composite or transition between the two. The dashed and continuous curves represent the \citet{Kauffmann2003} and \citet{kewley2006} criteria, respectively, that delimit the transition region. Each spaxel within the field of view corresponds to a point in this diagram. The dotted line, from \citet{cid10}, separates the regions occupied by LINERs and strong AGN (e.g. Seyfert nuclei), respectively. As the H$\beta$ emission is detected only within the inner $\sim$1 arcsec (2.11 kpc) in radius around the nucleus, the BPT diagram is limited to this  region. The line ratios from almost all spaxels included in the BPT diagram are located in the region occupied by strong AGN.  However, in regions farther away from the nucleus than $\sim$1 arcsec, the [N\, {\sc i}]$\lambda$6584/H$\alpha$ map shows values consistent with gas excitation by young stars, with log\,[N\, {\sc ii}]$\lambda$6584/H$\alpha<-0.2$.  
The bottom panels of Fig.\,\ref{fig:bpts} present the BPT diagram and flux line ratio maps for the broad component, showing line ratios typically observed in strong AGN.

\begin{figure*}
	\includegraphics[width=0.75\textwidth]{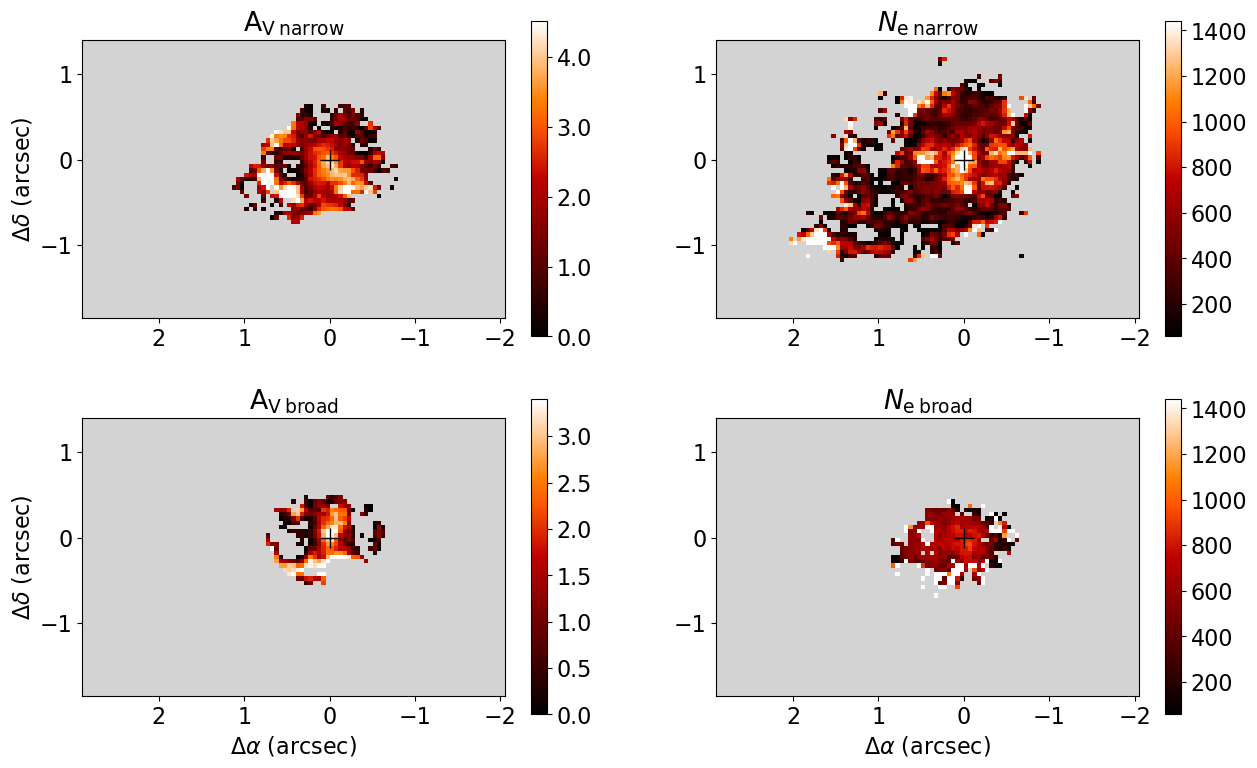}
    \caption{Visual extinction (left), derived from the H$\alpha$/H$\beta$ line ratio, and electron density (right) maps, obtained from the [S\,{\sc ii}]$\lambda\lambda$6717,6731 fluxes, for the narrow (top) and broad (bottom) components.  The color bars show the visual extinction in magnitudes and the electron density in cm$^{-3}$}.
    \label{fig:densidade}
\end{figure*}

In Figure~\ref{fig:densidade} we present the visual extinction (A$_V$; left panels) and electron density ($N_e$; right panels) maps for the narrow (top panels) and broad components (bottom panels). The A$_V$ values were estimated using the observed $\rm (H\alpha/H\beta)$ observed ratio, and the extinction law from \citet{ccm89}, following the procedure described in \citet{rogerio21}.  We were able to estimate A$_V$ only in the inner $\sim$1 arcsec because the H$\beta$ emission is not detected in locations further away from the nucleus. Typical A$_V$ values for both the broad and narrow components are in the range 1--3 mag. The mean values are 2.1$\pm$1.4 mag and 1.64$\pm$1.1 mag for the narrow and broad components, respectively. The quoted errors correspond to the standard deviation of the mean values. 

 The [N\,{\sc ii}]$\lambda$5755 line is detected in the nuclear spectrum, allowing us to use the [N\,{\sc ii}]$\lambda$6548,6583/[N\,{\sc ii}]$\lambda$5755 line ratio to estimate the electron temperature. Taking the line fluxes measured within a circular aperture of 0\farcs3 radius, roughly corresponding to the seeing of the GMOS observations, we find $T_e\approx 19\,000\pm$4\,000 K using the {\sc PyNeb} code. This temperature value is slightly larger than the values commonly reported for AGN photoionized gas \citep[$T_e\approx15\,000$\,K; ][]{dors17,dors20,revalski18,revalski21,rogemar21_teflut}, but still lower than shock dominated regions \citep[$T_e\gtrsim25\,000$\,K; ][]{rogemar21_te}. We estimate $N_e$ from the $[\ion{S}{ii}]\lambda6717/\lambda 6731$  line ratio using the  {\sc PyNeb} code \citep{luridiana15}, adopting an electron temperature of 19\,000 K. The $N_e$ maps for the narrow and broad components are shown in the top and bottom right panels of Fig.~\ref{fig:densidade}, respectively.  Typical $N_e$ values for both the broad and narrow components fall in the range from $\sim$200 to $\sim1300$ cm$^{-3}$ with the highest values observed within the inner 1 arcsec in radius from the nucleus, while the lowest values are seen mostly at greater distances. The mean values for the narrow and broad components are 580$\pm$370 and 1110$\pm$520 cm$^{-3}$, respectively.

\section{Discussion}

\subsection{Gas kinematics and outflows}
 The emission-line profiles in  IRAS19154 are well reproduced by two Gaussian functions. And in this case, we assume that the broad and narrow components are measuring properties of different groups of clouds. In this section we discuss the main characteristics of the regions emitting these components.

\subsubsection{The narrow component}

The gas velocity fields for the narrow components show mostly redshifts to the southeast and blueshifts to the northwest at least for distances larger than $\sim$1 arcsec -- that is, distances greater than 2.1 kpc at the galaxy. This behaviour suggests that the narrow component velocities, at least from these regions, are consistent with rotation in the galaxy plane. Within the inner 2.1 kpc, distortions from pure rotation are observed, as a twist in the zero velocity curve seen in H$\alpha$ and the blueshifts observed in the [O\,{\sc iii}]$\lambda$5007 velocity field.

 To characterize the velocity field, we employ an analytical model, assuming that the  gas follows circular orbits within the galactic plane, where the circular velocity is described by \citep{kruit78,bertola91},
\[ 
 V_{mod}(R,\psi)=V_{s}+ 
\]
\begin{equation} \label{eq-bertola}
    \frac{AR\cos(\psi-\psi_{0})\sin(i){\cos^{p}(i)}}{\{R^{2}[\sin^{2}(\psi-\psi_{0})+\cos^{2}(i)\cos^{2}(\psi-\psi_{0})]+{c_{0}}^{2}\cos^{2}(i)\}^{\frac{p}{2}}}, 
    \end{equation}
\noindent where $R$ is the distance from the nucleus projected on the plane of the sky and $\psi$ is its corresponding position angle; $V_{s}$ is the systemic velocity of the galaxy; $A$ is the velocity amplitude; 
$i$ is the inclination of the disc relative to the plane of the sky; $\Psi_0$ is the orientation of the line of nodes, measured North through East;  $c_{0}$ is a concentration parameter, defined as the radius where the rotation curve reaches 70\% of the velocity amplitude; and the parameter $p$ measures the slope of the rotation curve beyond the maximum amplitude.  

\begin{figure*}
	\includegraphics[width=0.98\textwidth]{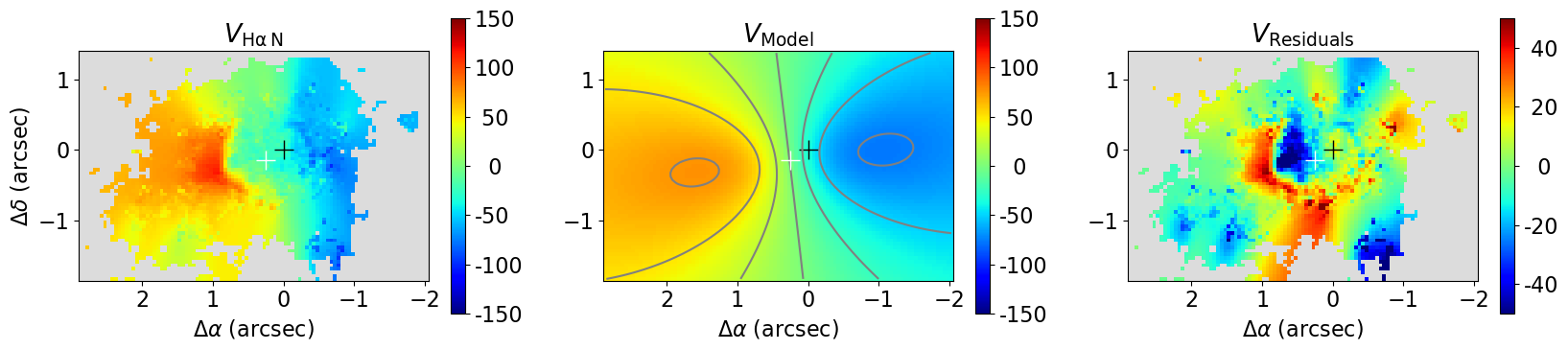}
    \caption{Observed H$\alpha$ velocity field for the narrow component (left), rotating disc model (centre) and residual map (right), defined as the difference between the observed velocities and the model. The black crosses mark the position of the nucleus, while the white crosses show the location of the kinematical centre.}
    \label{fig:bertola}
\end{figure*}

 Figure~\ref{fig:bertola} shows the observed velocity field for the H$_\alpha$ narrow component in the left panel, the best-fit model in the central panel,  and the residual map obtained by subtracting the model from the observed velocity field is shown in the right panel. At most locations, the residual velocities are smaller than $\sim$40\,km\,s$^{-1}$, indicating that the majority of the narrow component emitting gas is is rotating in the galaxy plane.  The kinematical centre located at ($\Delta\alpha$, $\Delta\delta$)=(0\farcs25, 0\farcs15) is slightly displaced from the position of the continuum peak. The disc kinematic major axis is oriented along the position angle $\Psi_0\approx96^\circ$ and the derived systemic velocity is $\approx$29975 km\,s$^{-1}$.   The highest velocity residuals are mostly observed within the inner 1 arcsec, where distinct behaviours are observed for the gas velocity field of the high ionization (traced by the [O\,{\sc iii}]$\lambda$5007 emission) and the low ionization (traced by H$\alpha$ and other emission lines) gas. The highest ionization gas is blueshifted by about 100 km\,s$^{-1}$, while the low ionization gas presents velocities close to zero in the central region. This result suggests that part of the gas emitting the narrow component is also outflowing, as is the case for the gas emitting the broad component, discussed in the next section. A puzzling velocity residual pattern is observed approximately 1 arcsec (2.1 kpc)  east/southeast of the nucleus, suggesting a compact rotation pattern,  that may be related with the interaction being suffered by this galaxy.

 The velocity dispersion values for the narrow component (Figure \ref{fig:gmos}) are in the range from 50 to 350\,km\,s$^{-1}$, with the highest values  being observed for the highest ionization gas within the inner arcsec and co-spatial with the locations where the [O\,{\sc iii}] velocity field shows blueshifts. This is also co-spatial with the region where the broad component is detected. This indicates that the distortions relative to pure rotation in the inner region of  IRAS19154 may be due to a disturbance of the gas in the plane of the disc by an outflow, traced by the broad component.

\subsubsection{The broad component}

 The velocity fields derived from the broad component show essentially blueshifts for all emission lines, with values reaching about $-$500\,km\,s$^{-1}$. These features are observed within the inner 2.1 kpc and are consistent with outflows. All velocity values are negative, indicating that the gas is mostly coming towards us. AGN outflows in ionized gas are usually bi-polar, but we do not see the redshifted counterpart of the blueshifted outflow. We attribute this result as due to the fact that the redshifted outflow is hidden behind the galaxy plane due extinction by dust.

The observed velocity fields can thus be understood as follows.
The broad component -- which is strongly blueshifted, represents a strong disturbance in the gas kinematics that 
can be attributed to the presence of an AGN driven wind. The narrow component could instead originate in gas clouds that are less affected by the presence of the nuclear outflow. The narrow component may be tracing the emission of the gas in the galaxy potential which is not strongly affected by the outflow in regions away from nucleus, while in the central arcsec the passage of the wind increases the turbulence of the gas in the host galaxy. Under the assumption that the broad component is due to an AGN driven outflow, we can now estimate is properties.

\subsubsection{Outflow properties}

We can estimate the mass outflow rate as being the ratio of the mass of the outflowing gas to the dynamical time for the gas to reach its present position. The mass of outflowing gas can be estimated by

\begin{equation}
     M_{g}=N_{e}m_{p}Vf  
\end{equation}
where $N_{e}$ is the electron density, $m_{p}$ is the mass of the proton, $V$ is the volume of the region where the outflow is observed and $f$ is the filling factor that can be estimated from

\begin{equation}
    L_{H\alpha}\approx f N_{e}^{2} j_{H\alpha}(T)V,
    \end{equation}
where j$_{H\alpha}$= 3.3534$\times$10$^{-25}$ erg cm$^{-3}$s$^{-1}$ \citep{Osterbrock2006} and $L_{H\alpha}$ is the H$\alpha$ luminosity emitted within the volume $V$.

Combining the above equations, we obtain 

\begin{equation}
M_g=\frac{m_p L_{H\alpha}}{N_e jH_\alpha (T).}
\end{equation}


 We estimate the observed H$\alpha$ luminosity of the outflow component by summing up the H$\alpha$ fluxes of the broad, blueshifted component of all spaxels and adopting a luminosity distance to IRAS19154 of 436\,Mpc.  We obtain $L_{{\rm H\alpha\,obs}}$=(8.9$\pm$0.4)$\times$10$^{41}$ erg s$^{-1}$. Using the extinction law from \citet{ccm89} and the mean visual extinction of $A_{\rm V\:out}=1.6\pm0.7$ for the outflow (Fig.~\ref{fig:densidade}), the resulting extinction corrected H$\alpha$ luminosity of the outflow is $L_{H\alpha}$=(3.0$\pm$1.4$\times$10$^{42}$\,erg\,s$^{-1}$. To estimate the mass of ionized hydrogen, we use an average density for the outflowing gas of $N_e\approx1110$\:cm$^{-3}$ obtained from the [S\:{\sc ii}] doublet (Fig.~\ref{fig:densidade}). The resulting mass is $M_g=(6.7\pm3.1)\times10^6$\:M$_\odot$.

 To estimate the dynamical time, we assume a radius of 0\farcs4 ($\approx850$~pc), which corresponds to half of the full width at half maximum of the flux distribution of the broad component of [O\,{\sc iii}]$\lambda$5007, and an outflow velocity of $V_{\rm out}\approx$500 km\:s$^{-1}$ obtained from the centroid velocity of the broad emission-line component (Fig.~\ref{fig:gmos}). Under these assumptions, the resulting mass-outflow rate is $\dot{M}_{\rm out}=4.0\pm2.6$\,M$_\odot$\,yr$^{-1}$, which is consistent with the values obtained for AGN with  luminosities similar to that of IRAS19154 \citep[e.g.][]{kakkad22, alice22, rogemar_nifs}.

We now compare the mass outflow rate with the mass accretion rate needed to feed the SMBH, which can be estimated as 
\begin{equation}
    \dot{M}_{\rm acc}=\frac{L_{\rm bol}}{\eta c^2},
\end{equation}
where $L_{\rm bol}$ is the bolometric luminosity and $c$ is the light speed. We estimate the bolometric luminosity from the observed [O\,{\sc iii}]$\lambda$5007 luminosity of the broad component, $L_{\rm [O\,III]}\approx2.2\times10^{42}$\,erg s$^{-1}$, using the relation $L_{\rm bol}=3500 L_{\rm [O\,III]}$ \citep{Heckman04}. This yields $L_{\rm bol} \approx7.8\times10^{45}$ erg\,s$^{-1}$. Assuming the radiative efficiency $\eta=0.1$, we derive a mass accretion rate of 1.4 M$_\odot$ yr$^{-1}$. Thus, the mass outflow rate in ionized gas is roughly 3 times larger than the accretion rate necessary to fuel the AGN.

The kinetic power of the ionized gas outflow can be estimated by
\begin{equation}
\dot{E}_{\rm out}=\frac{1}{2}\dot{M}(V_{\rm out}^2+3\sigma_{\rm out}^2)
\end{equation}
where $V_{\rm out}$ and $\sigma_{\rm out}$ are the outflow velocity and its velocity dispersion.  Using $V_{\rm out}\approx$500 km\:s$^{-1}$  and $\sigma_{\rm out}\approx$750 km\:s$^{-1}$ (Fig.~\ref{fig:gmos})  the resulting kinetic power of the outflow is $\dot{E}_{\rm out} = (2.5\pm1.6)\times10^{42}$\:erg\,s$^{-1}$.

Comparing the kinetic power with the bolometric luminosity, we find that $\dot{E}_{out}/L_{bol}$ is about 3$\times$10$^{-4}$. This value is lower than the outflow coupling efficiencies required by cosmological simulations for AGN feedback to become effective in suppressing star formation in the host galaxy \citep{Hopkins2010,Harrison2018}. However, one may consider that this kinetic efficiency corresponds only to the mechanical effect of the outflow seen in a specific gas phase and not to the total AGN energy.  Indeed, recent theoretical results indicate that low-power outflows can be efficient in quenching star formation in the host galaxy, if sustained for at least $\sim$1 Myr \citep{almeida23}.  

We note that, at a mass-outflow rate of 4M$_\odot$\,yr$^{-1}$, it will take only $\approx$0.25\,Myr for all the ionized gas mass emitting the broad component to be expelled. If the outflow duration is, for example, 10\,Myr \citep[assuming that the outflow is active for $\approx$10\% of a typical 100\,Myr AGN lifecycle; ][]{Schawinski15}, $\approx$\,10$^{8}$\:M$_\odot$ would be ejected during the activity cycle, in ionized gas alone.

\subsection{Scenario and comparison with previous studies}

The HST images in Fig.\,\ref{fig:HST} show that IRAS\,19154 is very irregular, with clear signatures of interaction, although no separate companion is seen in the images, indicating that the interaction is in an advanced stage. This interaction is the most probable trigger of the AGN.  This galaxy is part of a larger sample of interacting galaxies, selected from \citet{Darling2000}, for which we are conducting detailed studies based on observations at multiple wavelengths. All the galaxies included in the sample present OH megamaser emission, except for IRAS\,19154.
 Four out of the five galaxies studied previously exhibit clearer signs of interaction, when one can see the two components, and the presence of AGNs (except for IRAS\,17526+3253). However, IRAS\,19154 is the first object to show, besides the gas excitation by an AGN,  clear high-velocity outflows. It is possible that the  outflows begin when the interaction is in an advanced stage, but a larger sample will be needed to investigate this question.

In order to compare our results for IRAS\,1915 with those for our other studies of OHM galaxies, we now summarize our previous results. In \citet{Dinalva2015}, a multiwavelength analysis of IRAS\,16399-0937 revealed an embedded AGN in the northwestern nucleus of the galaxy. Another AGN was found in the galaxy IRAS\,23199+0123 which is a clear interacting pair of galaxies, a study in which we reported a new OH maser detection and related the maser sources to shocks driven by the AGN outflows \citep{Heka2018a}. Our study of IRAS\,03056+2034 revealed the presence of a circumnuclear star formation ring, in the center of which the BPT diagram and VLA data suggest the presence of another embedded AGN that is responsible for the gas excitation together with the star forming regions \citep{Heka2018b}. IRAS\,11506-3851 shows star-forming regions surrounding the nucleus but the multi wavelength analysis also suggests the presence of an embedded AGN \citep{Heka2020}. The only OHM galaxy for which we did not find evidence for the presence of an AGN is IRAS\,17526+3253, which hosts a H$_{2}$O maser and shows only excitation by star-forming regions, although a more comprehensive study may be necessary to confirm this result \citep{Dinalva2019}.

The study of the above OHM galaxies shows a repeated pattern where star forming regions surround the nucleus, which itself shows evidence of the presence of an AGN. In addition, all these galaxies present signatures of interactions in the recent past or are clearly in a merger process. Three of these galaxies are mergers and three of them are the result of a recent merger event and for most of them the data suggest the presence of an embedded AGN.

We note that, although our sample of studied galaxies is still small, we tentatively identify a trend where the AGN is more evident in galaxies that are in a more advanced merger stage, as well as its impact in the host galaxy, such as the outflows seen in IRAS\,1915. And in these cases the contribution of the AGN to the gas excitation is increased relative to that of the starburst activity. In the particular case of IRAS\,19154 (and others in an advanced merger stage), it seems that the AGN has already become powerful enough to produce outflows in ionized gas; in this environment, it is less likely that new stars will form close to the nucleus, but they may be observed farther out.

\section{Conclusions}

We have presented a two-dimensional study of the inner 6\,kpc$\times$4\,kpc region of the OH absorber galaxy IRAS\,19154 using GMOS - IFU observations. Our aim was to map its gas distribution and kinematics, as well as to determine its excitation mechanism. We summarize our main conclusions as follows:

\begin{itemize}

\item  Within the inner $\approx$\,2\,kpc radius, the gas presents two kinematic components:  
a broad component with velocity dispersion 500$\lesssim\sigma\lesssim900$\,km\,s$^{-1}$ and a narrow component with 50$\le \sigma\le 350$\,km\,s$^{-1}$, while in the outer regions, only the narrow component is observed;

\item We attribute the broad component to an AGN driven outflow, with velocities ranging from  $-$350 to  $-$500\,km\,s$^{-1}$. The outflow component is observed only in blueshifts, while the redshifted counterpart is  probably obscured by the disc of the galaxy;

\item  We attribute the narrow component to gas orbiting in the galaxy potential. It's velocity field is well reproduced by a rotating disk model at distances larger than $\sim2$kpc from the nucleus. A mild increase in the velocity dispersion and a higher discrepancy between the rotating disk model and the observed velocities are seen in the
central region, co-spatial with the detection of the outflow component. This suggests that he gas in the disc is disturbed either by the outflows or the interaction that this galaxy is undergoing;

\item The analysis of the emission line ratios reveals a powerful AGN ionizing and exciting the gas in the outflow, as well as the gas in the disc within the inner $\sim$2 kpc radius. Regions of low [N\,{\sc ii}]$\lambda$6584/H$\alpha$ are seen farther away, indicating gas emission associated with star forming regions. This is further supported by the detection of emission knots in the HST images in the continuum and H$\alpha$, beyond this region;

\item The mass of ionized gas in  the outflow is $\approx6.7\times10^6$\:M$_\odot$, and the corresponding mass-outflow rate is $\dot{M}_{\rm out}\approx4$\,M$_\odot$\,yr$^{-1}$. We estimate a kinetic power of the outflow of $2.5\times10^{42}$\:erg\,s$^{-1}$, which is nevertheless only $3\times10^{-4}$ L$_{\rm bol}$.

\item Although the above value is lower than prescribed by models for a significant impact on the host galaxy, at a mass-outflow rate of 4 M$_\odot$\,y$^{-1}$, all the present outflowing gas mass will be ejected from the central region in $\sim$0.25 Myr; $\approx$10$^8$ M$_\odot$ will be ejected in a $\approx$10\,Myr AGN lifecycle, in ionised gas alone.

\end{itemize}
The scenario we propose for IRAS\,19154 is that a recent interaction 
 has triggered the AGN and its outflow. IRAS\,19154 is the 6$^{th}$ galaxy we have observed with GMOS-IFU and the 5$^{th}$ for which we have found the presence of an AGN. Unlike the previous cases, it is an OH absorber, while the others are OHM galaxies. It is also the first showing the clear presence of an AGN outflow. This supports a trend of stronger AGN signatures in more advanced stages of interactions and mergers.

\section*{Acknowledgements}
We thank an anonymous reviewer for providing valuable comments that helped us improve this paper. This work is based on observations obtained at the Gemini Observatory, which is operated by the Association of Universities for Research in Astronomy, Inc., under a cooperative agreement with the NSF on behalf of the Gemini partnership: the National Science Foundation (United States), National Research Council (Canada), CONICYT (Chile), Ministerio de Ciencia, Tecnolog\'{i}a e Innovaci\'{o}n Productiva (Argentina), Minist\'{e}rio da Ci\^{e}ncia, Tecnologia e Inova\c{c}\~{a}o (Brazil), and Korea Astronomy and Space Science Institute (Republic of Korea).  This research has made use of NASA's Astrophysics Data System Bibliographic Services. This research has made use of the NASA/IPAC Extragalactic Database (NED), which is operated by the Jet Propulsion Laboratory, California Institute of Technology, under contract with the National Aeronautics and Space Administration. Support for programme HST-SNAP 11604 was provided by NASA through a grant from the Space Telescope Science Institute, which is operated by the Association of Universities for Research in Astronomy, Inc., under NASA contract NAS 5- 26555. 
 C.H. thanks CAPES for financial support. R.A.R. acknowledges the support from Conselho Nacional de Desenvolvimento Cient\'ifico e Tecnol\'ogico (CNPq; Proj. 400944/2023-5 \& 404238/2021-1)  and Funda\c c\~ao de Amparo \`a pesquisa do Estado do Rio Grande do Sul (FAPERGS; Proj. 21/2551-0002018-0). D.A.S. also CNPq, FAPERGS and CAPES for financial support. P.K acknowledges the support of the Department of Atomic Energy, Government of India, under the project 12-R\& D-TFR-5.02-0700.

\section*{Data Availability}

The GEMINI data used in this work is publicly available online via the GEMINI archive https://archive.gemini.edu/searchform/, with project code GN-2017B-Q-38. The VLA data is available at https://science.nrao.edu/facilities/vla/archive. Finally, the HST data is available at https://archive.stsci.edu/hst/ with project code 11604. The data cubes produced from these data can be shared on reasonable request to the corresponding author.



\bibliographystyle{mnras}

\begin{thebibliography}{99}

\bibitem[\protect\citeauthoryear{Allington-Smith et al.}{2002}]{allington-smith02} Allington-Smith, J.et al. 2002, PASP, 114, 892.

\bibitem[\protect\citeauthoryear{Allen et al.}{2008}]{Allen+08} Allen, M. G., Groves, B. A., Dopita, M. A., Sutherland, R. S. and Kewley, L. J., 2008, ApJS, 178, 20 

\bibitem[\protect\citeauthoryear{Almeida et al.}{2023}]{almeida23} Almeida, I., Nemmen, R., Riffel, R. A., 2023, MNRAS, 526, 217.

\bibitem[\protect\citeauthoryear{Arauujo et al.}{2023}]{araujo23} Araujo, B. L. C., Storchi-Bergmann, T.,  Rembold, S. B., Kaipper, A. L. P.,  Dall'Agnol de Oliveira, B.,  2023, MNRAS, 522, 5165.

\bibitem[\protect\citeauthoryear{Bae \& Woo}{2016}]{bae16} Bae, H.-J., Woo, J.-H., 2016, ApJ, 828, 97.

\bibitem[\protect\citeauthoryear{Baldwin, Phillips \& Terlevich}{1981}]{baldwin81} Baldwin, J. A., Phillips, M. M., Terlevich, R., 1981, PASP, 93, 5

\bibitem[\protect\citeauthoryear{Barnes \& Hernquist}{1992}]{Barnes92} Barnes, J. E., \& Hernquist, L. 1992, ARA\&A, 30, 705

\bibitem[\protect\citeauthoryear{Berton et al.}{2018}]{Berton2018}
Berton, M, 2018, A\&A, 614, 87.


\bibitem[\protect\citeauthoryear{Bertola et al.}{1991}]{bertola91} Bertola, F., Bettoni, D., Danziger, J., Sadler, E., Sparke, L., de Zeeuw, T., 1991, ApJ, 373, 369


\bibitem[\protect\citeauthoryear{Brum et al.}{2019}]{brum19} Brum, C. et al., 2019, MNRAS, 486, 691


\bibitem[\protect\citeauthoryear{Cardelli, Clayton \& Mathis}{1989}]{ccm89} Cardelli, J. A., Clayton, G. C., Mathis, J. S., 1989, ApJ, 345, 245.

\bibitem[\protect\citeauthoryear{Cid Fernandes et al.}{2010}]{cid10} Cid Fernandes, R., Stasi\'nska, G., Schlickmann, M. S., Mateus, A., Vale Asari, N., Schoenell, W., Sodr\'e, L., 2010, MNRAS, 403, 1036.

\bibitem[\protect\citeauthoryear{Condon et al.}{1998}]{Condon1998} Condon, J. J., Cotton, W. D., Greisen, E. W., Yin, Q. F., Perley, R. A., Taylor, G. B., Broderick, J. J., 1998, AJ, 115, 1693.

\bibitem[\protect\citeauthoryear{Dall'Agnol de Oliveira et al.}{2021}]{bruno21} 
Dall'Agnol de Oliveira, B. et al., 2021, MNRAS, 504, 3890.

\bibitem[\protect\citeauthoryear{Darling \& Giovanelli}{2000}]{Darling2000}
Darling J., Giovanelli R., 2000, ApJ, 119, 3003

\bibitem[\protect\citeauthoryear{Darling \& Giovanelli}{2001}]{Darling2001}
Darling J., Giovanelli R., 2001, ApJ, 121, 1278

\bibitem[\protect\citeauthoryear{Darling \& Giovanelli}{2002}]{Darling2002}
Darling J., Giovanelli R., 2002, ApJ, 124, 100

\bibitem[\protect\citeauthoryear{Deconto-Machado et al.}{2022}]{alice22} Deconto-Machado, A. et al., 2022, A\&A, 659, 131.

\bibitem[\protect\citeauthoryear{do Nascimento et al.}{2022}]{doNascimento22} do Nascimento, J. C. et al, 2022, 513, 807.


\bibitem[\protect\citeauthoryear{Dors et al.}{2017}]{dors17} Dors, O. L.,  Arellano-C\'ordova, K. Z., Cardaci, M. V., H\"agele, G. F., 2017, MNRAS, 468, 113.

\bibitem[\protect\citeauthoryear{Dors et al.}{2020}]{dors20} Dors, O. L., Maiolino, R., Cardaci, M. V., H\"agele, G. F., Krabbe, A. C., P\'erez-Montero, E.,  Armah, M., 2020, MNRAS, 496, 3209.

\bibitem[\protect\citeauthoryear{Freitas et al.}{2018}]{freitas18} Freitas, I. C. et al., 2018, MNRAS, 476, 2760.

\bibitem[\protect\citeauthoryear{Haan et al}{2011}]{Haan2011} Haan, S., Surace, J. A., Armus, L., et al. 2011, AJ, 141, 100

\bibitem[\protect\citeauthoryear{Harrison et al.}{2018}]{Harrison2018} Harrison C. M., Costa T., Tadhunter C. N., Flütsch A., Kakkad D., Perna M.,Vietri G., 2018, Nature Astronomy, 2, 198

\bibitem[\protect\citeauthoryear{Heckman et al}{2004}]{Heckman04} Heckman T. M., Kauffmann G., Brinchmann J., Charlot S., Tremonti C., White S. D. M., 2004, ApJ, 613, 109.

\bibitem[\protect\citeauthoryear{Hekatelyne et al}{2018a}]{Heka2018a} 
Hekatelyne, C., Riffel, R. A., Sales, D., Robinson, A., Gallimore, J., Storchi-Bergmann, T.,Kharb, P., O'Dea, C. Baum, S., 2018a MNRAS, v. 474, p. 5319

\bibitem[\protect\citeauthoryear{Hekatelyne et al}{2018b}]{Heka2018b}
Hekatelyne, C., Riffel, R. A., Sales, D., Robinson, A., Storchi-Bergmann, T., Kharb,P., Gallimore, J., Baum, S., O'Dea, C., 2018b, MNRAS, v. 479, p 3966

\bibitem[\protect\citeauthoryear{Hekatelyne et al}{2020}]{Heka2020}
Hekatelyne, C., Riffel, R. A., Storchi-Bergmann, T., Kharb,P., Robinson, A., Sales, D., Cassanta, C., 2020, MNRAS,

\bibitem[\protect\citeauthoryear{Hoopes et al.}{1999}]{Hoopes99} Hoopes, C., G., Walterbos, R. A. M., Rand, R. J., 1999, ApJ, 552, 669.

\bibitem[\protect\citeauthoryear{Hook et al.}{2004}]{hook04} Hook, I., Jorgensen, I., Allington-Smith, J. R., Davies, R. L., Metcalfe, N., Murowinski, R. G., Crampton, D., 2004, PASP, 116, 425

\bibitem[\protect\citeauthoryear{Hopkins et al}{2006}]{Hopkins2006} Hopkins, P. F., Hernquist, L., Cox, T. J., et al. 2006, ApJS, 163, 1

\bibitem[\protect\citeauthoryear{Hopkins \& Elvis}{2010}]{Hopkins2010} Hopkins P. F., Elvis M., 2010, MNRAS, 401, 7


\bibitem[\protect\citeauthoryear{Kakkad et al.}{2022}]{kakkad22} Kakkad, D. et al., 2022, MNRAS, 511, 2105.

\bibitem[\protect\citeauthoryear{Kauffmann et al.}{2003}]{Kauffmann2003} Kauffmann, G. et al. 2003a, MNRAS, 346, 1055

\bibitem[\protect\citeauthoryear{Kewley et al.}{2006}]{kewley2006} Kewley, L. J, Groves, B, Kauffmann G., Heckman, T., 2016, MNRAS, 372


\bibitem[\protect\citeauthoryear{Krabbe et al.}{2014}]{krabbe14} Krabbe, A. C.  et al., 2014, MNRAS, 437, 1155.


\bibitem[\protect\citeauthoryear{Kraemer et al.}{1999}]{kraemer99} Kraemer, S. B., Ho, L. C., Crenshaw, D. M., Shields, J. C., Filippenko, A. V., 1999, ApJ, 520, 564.


\bibitem[\protect\citeauthoryear{Kukula et al.}{1998}]{Kukula1998} Kukula, M. J., Dunlop, J. S., Hughes, D. H., Rawlings, S., 1998, MNRAS, 297, 366


\bibitem[\protect\citeauthoryear{Lacy et al.}{2020}]{Lacy2020} Lacy, M., 2020, PASP, 132c5001L


\bibitem[\protect\citeauthoryear{Lo}{2005}]{Lo2005} Lo K. Y., 2005, ARA\&A, 43, 625

\bibitem[\protect\citeauthoryear{Ludwig et al.}{2012}]{ludwig12} Ludwig, R. R., Greene, J. E., Barth, A. J., Ho, L. C., 2012, ApJ, 756, 51

\bibitem[\protect\citeauthoryear{Luridiana, Morisset \& Shaw}{2015}]{luridiana15} Luridiana, V., Morisset, C., Shaw, R. A.,
 2015, A\&A, 573, A42

\bibitem[\protect\citeauthoryear{Mihos}{1995}]{Mihos95} Mihos, J. C. 1995, ApJ, 438, 75.


\bibitem[\protect\citeauthoryear{Nakanishi}{1997}]{Nakanishi1997} Nakanishi, K., Takata, T., Yamada, T., Takeuchi, T. T., Shiroya, R., Miyazawa, M., Watanabe, S., \& Saito, M. 1997, ApJS, 112, 245

\bibitem[\protect\citeauthoryear{Osterbrock \& Ferland}{2006}]{Osterbrock2006} Osterbrock, D. E., Ferland, G. J., 2006, Sausalito, CA: University Science Books.

\bibitem[\protect\citeauthoryear{Perez et al.}{2011}]{Perez11} Perez, J. Michel-Dansac, L. Tissera, P. B., 2011, MNRAS, 417, 580.

\bibitem[\protect\citeauthoryear{Petric et al.}{2018}]{Petric18} Petric, A. O. et al., 2018, AJ, 156, 295.

\bibitem[\protect\citeauthoryear{Revalski et al.}{2018}]{revalski18} Revalski, M., Crenshaw, D. M., Kraemer, S. B., Fischer, T. C., Schmitt, H. R., Machuca, C., 2018, ApJ, 856, 46.
 
 \bibitem[\protect\citeauthoryear{Revalski et al.}{2021}]{revalski21} Revalski, M. et al., 2021, ApJ, 910, 139.
 
\bibitem[\protect\citeauthoryear{Riffel et al.}{2021a}]{rogemar21_te} Riffel, R. A. et al., 2021a, MNRASL, 501, L54.

\bibitem[\protect\citeauthoryear{Riffel et al.}{2021b}]{rogemar21_teflut} Riffel, R. A., Dors, O. L.,Krabbe, A. C., Esteban, C., 2021b, MNRASL, 506, L11.

\bibitem[\protect\citeauthoryear{Riffel et al.}{2023}]{rogemar_nifs} Riffel, R. A. et al., 2023, MNRAS, 521, 1832.



\bibitem[\protect\citeauthoryear{Riffel et al.}{2021c}]{rogerio21} Riffel, R. et al., 2021, MNRAS, 501, 4064.

\bibitem[\protect\citeauthoryear{Rossa \& Dettmar}{2000}]{Rossa00} Rossa, J.; Dettmar, R. -J, A\&A, 359, 433.

\bibitem[\protect\citeauthoryear{Rossa \& Dettmar}{2003}]{Rossa03} Rossa, J.; Dettmar, R. -J, A\&A, 406, 505.

\bibitem[\protect\citeauthoryear{Ruschel-Dutra}{2020}]{Ruschel-Dutra2020} Ruschel-Dutra D., 2020, danielrd6/ifscube v1.0,doi:10.5281/zenodo.3945237

\bibitem[\protect\citeauthoryear{Ruschel-Dutra et al.}{2021}]{Ruschel-Dutra2021} Ruschel-Dutra D. et al., 2021, MNRAS, 507, 74.


\bibitem[\protect\citeauthoryear{Schawinski et al.}{2015}]{Schawinski15} Schawinski, K., Koss, M.,  Berney, S., Sartori, L. F., 2015, MNRAS, 451, 2517.

\bibitem[\protect\citeauthoryear{Schwarzkopf \& Dettmar}{2000}]{Schwarzkopf00} Schwarzkopf, U., Dettmar, R. -J., 2000, A\&A, 361, 451.

\bibitem[\protect\citeauthoryear{Sales et al}{2015}]{Dinalva2015} 
Sales, D. A., Robinson, A., Axon, D. J., et al., 2015, ApJ, 799, 25

\bibitem[\protect\citeauthoryear{Sales et al}{2019}]{Dinalva2019} 
Sales, D., Robinson, A., Riffel, R. A, Storchi-Bergmann, T, Gallimore, J F, Kharb, P, Baum, S, O'dea, C, Hekatelyne, C, Ferrari, F., 2019, MNRAS, v. 486, 3350

\bibitem[\protect\citeauthoryear{Sanders et al}{1988}]{sanders88} Sanders, D. B., Soifer, B. T., Elias, J. H., et al. 1988, ApJ, 325, 74

\bibitem[\protect\citeauthoryear{Soifer et al}{1987}]{Soifer87} Soifer, B. T. et al. 1987, ApJ, 320, 238.

\bibitem[\protect\citeauthoryear{Storchi-Bergmann \& Schnorr-M\"uller}{2019}]{SB-SM19} Storchi-Bergmann, T. \& Schnorr-M\"uller, A 2019, Nature Astronomy, 3, 48

\bibitem[\protect\citeauthoryear{Tissera et al.}{2002}]{Tissera02} Tissera, P. B., Dom\'inguez-Tenreiro, R., Scannapieco, C., S\'aiz, A., 2002, MNRAS, 333, 327.

\bibitem[\protect\citeauthoryear{Tody}{1986}]{tody86} Tody, D. 1986, The IRAF Data Reduction and Analysis System in Proc. SPIE  Instrumentation in Astronomy VI, ed. D.L. Crawford, 627, 733  

\bibitem[\protect\citeauthoryear{Tody}{1993}]{tody93} Tody, D. 1993, IRAF in the Nineties" in Astronomical Data Analysis Software and Systems II, A.S.P. Conference Ser., Vol 52, eds. R.J. Hanisch, R.J.V. Brissenden, J. Barnes, 173.

\bibitem[\protect\citeauthoryear{Ulvestad et al.}{2005}]{Ulvestad2005} Ulvestad, J. S., Wong, D. S., Taylor, G. B., Gallimore, J. F., Mundell, C. G., 2005, ApJ, 130, 936,

\bibitem[\protect\citeauthoryear{van der Kruit \& Allen}{1978}]{kruit78} van der Kruit, P.C., Allen, R.J., 1978, ARA\&A, 16, 103


\bibitem[\protect\citeauthoryear{van Dokkum}{2001}]{lacos} Van Dokkum, P. G., 2001, PAP, 113, 789.


\end{thebibliography}






\bsp	
\label{lastpage}
\end{document}